\PassOptionsToPackage{unicode}{hyperref}
\PassOptionsToPackage{hyphens}{url}
\PassOptionsToPackage{dvipsnames,svgnames,x11names}{xcolor}
\documentclass[
]{article}

\usepackage{amsmath,amssymb}
\usepackage{iftex}
\ifPDFTeX
  \usepackage[T1]{fontenc}
  \usepackage[utf8]{inputenc}
  \usepackage{textcomp} % provide euro and other symbols
\else % if luatex or xetex
  \usepackage{unicode-math}
  \defaultfontfeatures{Scale=MatchLowercase}
  \defaultfontfeatures[\rmfamily]{Ligatures=TeX,Scale=1}
\fi
\usepackage{lmodern}
\ifPDFTeX\else  
\fi
\IfFileExists{upquote.sty}{\usepackage{upquote}}{}
\IfFileExists{microtype.sty}{% use microtype if available
  \usepackage[]{microtype}
  \UseMicrotypeSet[protrusion]{basicmath} % disable protrusion for tt fonts
}{}
\makeatletter
\@ifundefined{KOMAClassName}{% if non-KOMA class
  \IfFileExists{parskip.sty}{%
    \usepackage{parskip}
  }{% else
    \setlength{\parindent}{0pt}
    \setlength{\parskip}{6pt plus 2pt minus 1pt}}
}{% if KOMA class
  \KOMAoptions{parskip=half}}
\makeatother
\usepackage{xcolor}
\ifx\paragraph\undefined\else
  \let\oldparagraph\paragraph
  \renewcommand{\paragraph}[1]{\oldparagraph{#1}\mbox{}}
\fi
\ifx\subparagraph\undefined\else
  \let\oldsubparagraph\subparagraph
  \renewcommand{\subparagraph}[1]{\oldsubparagraph{#1}\mbox{}}
\fi

\providecommand{\tightlist}{%
  \setlength{\itemsep}{0pt}\setlength{\parskip}{0pt}}\usepackage{longtable,booktabs,array}
\usepackage{calc} % for calculating minipage widths
\usepackage{etoolbox}
\makeatletter
\patchcmd\longtable{\par}{\if@noskipsec\mbox{}\fi\par}{}{}
\makeatother
\IfFileExists{footnotehyper.sty}{\usepackage{footnotehyper}}{\usepackage{footnote}}
\makesavenoteenv{longtable}
\usepackage{graphicx}
\makeatletter
\def\maxwidth{\ifdim\Gin@nat@width>\linewidth\linewidth\else\Gin@nat@width\fi}
\def\maxheight{\ifdim\Gin@nat@height>\textheight\textheight\else\Gin@nat@height\fi}
\makeatother
\setkeys{Gin}{width=\maxwidth,height=\maxheight,keepaspectratio}
\makeatletter
\def\fps@figure{htbp}
\makeatother

\usepackage{arxiv}
\usepackage{orcidlink}
\usepackage{amsmath}
\usepackage[T1]{fontenc}
\makeatletter
\@ifpackageloaded{caption}{}{\usepackage{caption}}
\AtBeginDocument{%
\ifdefined\contentsname
  \renewcommand*\contentsname{Table of contents}
\else
  \newcommand\contentsname{Table of contents}
\fi
\ifdefined\listfigurename
  \renewcommand*\listfigurename{List of Figures}
\else
  \newcommand\listfigurename{List of Figures}
\fi
\ifdefined\listtablename
  \renewcommand*\listtablename{List of Tables}
\else
  \newcommand\listtablename{List of Tables}
\fi
\ifdefined\figurename
  \renewcommand*\figurename{Figure}
\else
  \newcommand\figurename{Figure}
\fi
\ifdefined\tablename
  \renewcommand*\tablename{Table}
\else
  \newcommand\tablename{Table}
\fi
}
\@ifpackageloaded{float}{}{\usepackage{float}}
\floatstyle{ruled}
\@ifundefined{c@chapter}{\newfloat{codelisting}{h}{lop}}{\newfloat{codelisting}{h}{lop}[chapter]}
\floatname{codelisting}{Listing}

\makeatother
\makeatletter
\@ifpackageloaded{caption}{}{\usepackage{caption}}
\@ifpackageloaded{subcaption}{}{\usepackage{subcaption}}
\makeatother
\makeatletter
\@ifpackageloaded{tcolorbox}{}{\usepackage[skins,breakable]{tcolorbox}}
\makeatother
\makeatletter
\@ifundefined{shadecolor}{\definecolor{shadecolor}{rgb}{.97, .97, .97}}
\makeatother
\ifLuaTeX
  \usepackage{selnolig}  % disable illegal ligatures
\fi
\IfFileExists{bookmark.sty}{\usepackage{bookmark}}{\usepackage{hyperref}}
\IfFileExists{xurl.sty}{\usepackage{xurl}}{} % add URL line breaks if available
\hypersetup{
  pdftitle={The Public Health and Environmental Surveillance Open Data Model (PHES-ODM) version 3: an open relational data model and interoperability framework for wastewater surveillance},
  pdfauthor={Mathew Thomson; Jean-David Therrien; Nikho Hizon; Janet Lin; Martin Wellman; Eugen-Sorin Sion; Carol Bennett; Peter Van Rolleghem; Douglas Manuel},
  pdfkeywords={wastewater-based epidemiology, wastewater
surveillance, open data model, relational database, data
interoperability, FAIR data, public health surveillance, data
dictionary, metadata standard},
  colorlinks=true,
  linkcolor={blue},
  filecolor={Maroon},
  citecolor={Blue},
  urlcolor={Blue},
  pdfcreator={LaTeX via pandoc}}

\newcommand{\runninghead}{A Preprint }
\title{The Public Health and Environmental Surveillance Open Data Model
(PHES-ODM) version 3: an open relational data model and interoperability
framework for wastewater surveillance\thanks{This work is licensed under
the Creative Commons Attribution 4.0 International License (CC BY
4.0).}}
\def\asep{\\\\\\ } % default: all authors on same column
\def\asep{\And }
\author{\textbf{Mathew
Thomson}~\orcidlink{0000-0002-8155-3840}\\\\Ottawa Hospital Research
Institute, University of Ottawa\\Ottawa, ON\\\asep\textbf{Jean-David
Therrien}~\orcidlink{0000-0002-8561-2916}\\\\Ghent University Faculty of
Bioscience Engineering\\Ghent\\\\ModelEAU, Université Laval\\Québec
City, QC\\\asep\textbf{Nikho Hizon}\\\\Public Health Agency of Canada
National Microbiology Laboratory\\Winnipeg, MB\\\asep\textbf{Janet
Lin}\\\\Public Health Agency of Canada National Microbiology
Laboratory\\Winnipeg, MB\\\asep\textbf{Martin Wellman}\\\\Ottawa
Hospital Research Institute, University of Ottawa\\Ottawa,
ON\\\asep\textbf{Eugen-Sorin Sion}\\\\European Commission, Joint
Research Centre\\Ispra\\\asep\textbf{Carol Bennett}\\\\Ottawa Hospital
Research Institute, University of Ottawa\\Ottawa, ON\\\asep\textbf{Peter
Van Rolleghem}~\orcidlink{0000-0003-1695-1313}\\\\ModelEAU, Université
Laval\\Québec City, QC\\\asep\textbf{Douglas
Manuel}~\orcidlink{0000-0003-0912-0845}\\\\Ottawa Hospital Research
Institute, University of Ottawa\\Ottawa,
ON\\\href{mailto:dmanuel@ohri.ca}{dmanuel@ohri.ca}}
\date{2026-04-08}
\begin{document}
\maketitle
\begin{abstract}
Wastewater surveillance (WWS) has emerged as a valuable tool for public
health surveillance, particularly since the COVID-19 pandemic. Its
long-term utility is constrained, however, by fragmented data systems,
inconsistent metadata practices, and poor interoperability. The Public
Health and Environmental Surveillance Open Data Model (PHES-ODM) was
developed as an open, collaborative framework to standardize WWS data
and support transparent, ethical data use aligned with FAIR principles.
Adopted by the Public Health Agency of Canada and adapted by the EU
Sewage Sentinel System, the model is now used in over 25 countries. This
paper introduces version 3 of the model, which addresses persistent
barriers to interoperability and data utility. Key enhancements include
new tables for public health actions, external repository linkages
(e.g., GISAID, GenBank), and analytical workflow documentation, as well
as support for complex relational linkages across sites, samples,
measures, and populations. Tools for mapping across other data formats,
including PHA4GE and the US CDC National Wastewater Surveillance System,
and for supporting long and wide data formats are also introduced. We
compare PHES-ODM against six other WWS data standards across 25
features. Balancing robustness with usability, PHES-ODM v3 provides a
scalable, modular infrastructure adaptable to diverse WWS and
environmental surveillance programs.
\end{abstract}
{\bfseries \emph Keywords}
\def\sep{\textbullet\ }
wastewater-based epidemiology \sep wastewater surveillance \sep open
data model \sep relational database \sep data interoperability \sep FAIR
data \sep public health surveillance \sep data dictionary \sep 
metadata standard

\ifdefined\Shaded\renewenvironment{Shaded}{\begin{tcolorbox}[boxrule=0pt, interior hidden, breakable, sharp corners, enhanced, borderline west={3pt}{0pt}{shadecolor}, frame hidden]}{\end{tcolorbox}}\fi

\hypertarget{introduction}{%
\section{1. Introduction}\label{introduction}}

Historically used to track fecal-transmitted and water-borne pathogens,
wastewater surveillance (WWS) of public health threats has become
increasingly discussed since the emergence of SARS-CoV-2 and the
COVID-19 pandemic {[}1-2{]}. Since that time almost 300 universities,
with over 4,500 sites in over 70 countries, implemented WWS programs
{[}3-4{]}. The World Health Organization, among other experts, have
lauded the rapid development and adoption of WWS as a valuable addition
to the existing pathogen surveillance programs, and celebrated its
utility as a valuable tool for promoting and protecting human health
{[}5-6{]}. As a part of the boom in WWS, the Rockefeller foundation
convened the Wastewater Action Group {[}7{]}, and the Bill and Melinda
Gates Foundation has made WWS part of its Enterics, Diagnostics,
Genomics \& Epidemiology (EDGE) program {[}8{]}. Voluntary and global
networks-of-networks, such as GLOWACON {[}9{]}, have also been
assembled, with an aim to establish global networks, and networks of
networks, to support WWS. These networks have succeeded in convening and
supporting the community of practice and have fostered knowledge-sharing
and the deployment of innovations for informed public health
decision-making {[}9{]}.

The rise of WWS, however, is not without challenges {[}7, 10{]}. In
particular, data from WWS is not made widely accessible, and this
restricted access to data hampers large-scale coordination and
meta-analytical research, ultimately obstructing the integration of
wastewater insights into actionable public health policy. Challenges
have been persistent since the inception of many programs {[}3-4{]}. As
such, addressing barriers and improving support for open data sharing is
a priority issue for global networks. Networks have begun to
collaboratively generate and release guidance documents on how to
communicate wastewater-related public health findings, as well as best
practices on how to perform the analyses required. Networks are also
discussing standardization, best practices in statistical modeling, and
feasibility studies on open-sharing platforms {[}7-9{]}. For new
surveillance systems like WWS, the absence of (or unwillingness to
adopt) existing data standards, insufficient system literacy, lack of
guidance on the interpretation of data, insufficient data
infrastructure, poorly understood or explained laboratory
infrastructure, and the diversity of fields involved in WWS in
particular, all pose important challenges to the longevity and utility
of these programs, as well as to data sharing and coordination {[}11{]}.
Furthermore, issues around consistent data reporting, the lack of
interoperable data formatting, inconsistent metadata collection, and
reticence toward data sharing all need to be addressed if WWS is to
advance and become a permanent fixture of the public health landscape
{[}11-12{]}.

The Public Health and Environmental Surveillance Open Data Model
(PHES-ODM) was developed during the inception of WWS for SARS-CoV-2 in
Ottawa, Canada, to standardise data reporting and storage {[}13-14{]}.
By structuring real-world entities and recording critical
metadata---such as collection methods and analytic protocols---the model
provided valuable context as diverse labs began reporting to the Ontario
provincial government. From the first prototypes, this metadata was
instrumental in distinguishing true infection trends from site-specific
technical variations. Originally developed in collaboration with the
Delatolla lab (Ottawa) and the CentrEAU research cluster (the province
of Québec), the PHES-ODM filled a void where no national or
international standards yet existed. The remarkable success and impact
of open data sharing from COVID-19 clinical surveillance systems
{[}15{]} and growing interest in PHES-ODM led to the adoption of a fully
open and collaborative platform. Major contributing organizations
included the European Commission Joint Research Centre, the United
States Centers for Disease Control and Prevention's (USCDC) National
Wastewater Surveillance System (NWSS){[}16{]}, and the Public Health
Agency of Canada {[}17{]}, along with Canadian provincial WWS programs
in Ontario and Québec. The model was also adapted by organizations in
the private sector, notably the CETo epidemiological software program
{[}18{]} and AdvanSentinel's program {[}19{]}. Collaboration and
participation continued to expand with regular working group meetings,
and membership from all world regions except Africa.

The PHES-ODM was designed as an open, collaborative data standard to
provide a common language when discussing WWS globally. It emerged to
fill the need for a standard or openly available structure for recording
and storing WWS data. This was an issue to be solved because
transparency in maintaining data is foundational to its ethical handling
{[}20{]}, and transparency and metadata are crucial to principles of
data justice. According to D'Ignazio and Klein {[}21{]}, ``consider
context'' is the sixth principle of data feminism; without
context---provided through metadata---data remains prone to
misinterpretation. When this data is about humans and human health,
misinterpretation can cause immense harm. Data is also only useful if it
is understandable and analyzable---without access and context, there is
no possibility to repurpose the data. Governments also risk squandering
public funds and eroding public trust if collected data cannot be used.
To better define guidelines to assess data transparency and usability,
the FAIR Data Principles assert that responsible and ethically managed
data should be Findable, Accessible, Interoperable, and Reusable
{[}12{]}. In line with these principles, we developed open data
dictionary to support metadata management. Data dictionaries, resources
that support data curation by providing standardised definitions of
terms for use in metadata, are foundational to transparent data
collection and management. They also facilitate ethical handling of
sensitive data, and support transparency in research and data inquiry
through the entire lifecycle of the data {[}20{]}. These kinds of good
data management and stewardship protocols are prerequisites to
advancement and innovation in any field, but particularly in developing
fields such as WWS and wastewater-based epidemiology (WBE) where sharing
and collaboration between areas of practice and institutions are
essential.

In its approach to WWS data, the PHES-ODM aims to emulate and adapt
approaches from similar systems within public health, like Logical
Observation Identifiers Names and Codes (LOINC), a universal standard
and database for medical laboratory measures {[}22{]}. In this way,
version 1 of the PHES-ODM enabled users to record the basic WWS data and
metadata in a standardised way, facilitating the exchange and
aggregation of these data to improve coordinated surveillance efforts
{[}13{]}. Beyond technical utility, the model was meant to improve
public health outcomes by enhancing wastewater and environmental
surveillance and epidemiology through interoperable, transparent, and
efficient data collection and use. By using a relational database
structure, the model links data across the entire analytic lifecycle.
With unique identifiers for sites, samples, measures, and other
attributes, the relational structure allows for each piece of the WWS
process to be recorded and accounted for, while avoiding issues like
data duplication, inconsistency, and challenges with insertion and
deletion {[}14, 23{]}. Version 1 focused on recording sample information
and measures, site information and measures, and clinical surveillance
information on COVID-19 infections. With version 2 of the PHES-ODM, the
model collapsed site and sample measures into a singular measures table,
while expanding out additional provisions for methods, protocols, sample
provenance, among many other fields {[}14{]}. While the original version
1 {[}13{]} could still be effective for small-scale SARS-CoV-2
surveillance programs, version 2 went much deeper to collect information
and respond to growing demands for additional metadata fields. It also
provided templates for reporting variants and mutations of SARS-CoV-2,
and the tracking and reporting of additional pathogens. The structure of
version 2 further built on the original relational format and expanded
linkages using keys. By integrating this relational approach with
entity-relationship modelling frameworks {[}24{]}, the PHES-ODM provides
a robust and highly customizable database structure that ensures data
provenance is explicit in the metadata.

While the adoption and use of both version 1 and 2 of the model have
been very successful, global progress on WWS programs is now at risk. As
political priorities shift toward managing polycrises {[}25{]} and an
endemic approach to SARS-CoV-2, there is a danger of losing the momentum
gained in global wastewater surveillance {[}6{]}. A concerted effort to
continue to support and develop WWS programs and initiatives is required
to maintain this system for the next pandemic. Initiatives to assess
programs are forthcoming {[}26{]}, but data models need to be responsive
to additional context and expanding programs. After the release and wide
global adoption of version 2 of the PHES-ODM {[}14{]}, it is now the
official model used by the Public Health Agency of Canada for their WWS
program {[}17{]} and has been adopted and modified by the European Union
Sewage Sentinel System for SARS-CoV-2 {[}27{]}, among other programs and
initiatives. WWS data is still, however, heavily siloed and in a crisis
of a lack of interoperability and sharing.

Building on the foundation of version 2, this paper introduces version 3
of the PHES-ODM. The revisions and expansions in this latest iteration
were developed in direct response to consultations with current users
and data dictionary developers. We explore the model's evolution,
focusing on its ability to balance robust functionality with ease of use
while supporting interoperability within complex epidemiological
environments. Ultimately, Version 3 offers structural solutions to
common data challenges in WWS; this paper details the model's
architecture, its target audience, implementation, and overall utility
for global WWS programs and as a solution to data challenges in the
larger field of WWS.

\hypertarget{materials-and-methods}{%
\section{2. Materials and Methods}\label{materials-and-methods}}

\hypertarget{overview-of-the-phes-odm-the-structure-of-the-model}{%
\subsection{2.1. Overview of the PHES-ODM: the structure of the
model}\label{overview-of-the-phes-odm-the-structure-of-the-model}}

A more detailed account of the underlying basic structure from version 2
of the PHES-ODM, and the rationale that led to this structure, has been
described elsewhere {[}14{]}. To review the general principles, however,
we need to acknowledge that wastewater and environmental surveillance
data are complex and that many factors, originating from vastly
different disciplines, must be considered when interpreting them.
Information about the geographic area, its population and composition;
environmental and industrial factors in and around the sampling site;
sample composition and treatment details; information on the methodology
and analytical assays employed; as well as data on measurement and
sample quality, are all crucial to interpreting WWS data.

\begin{figure}

{\centering \includegraphics{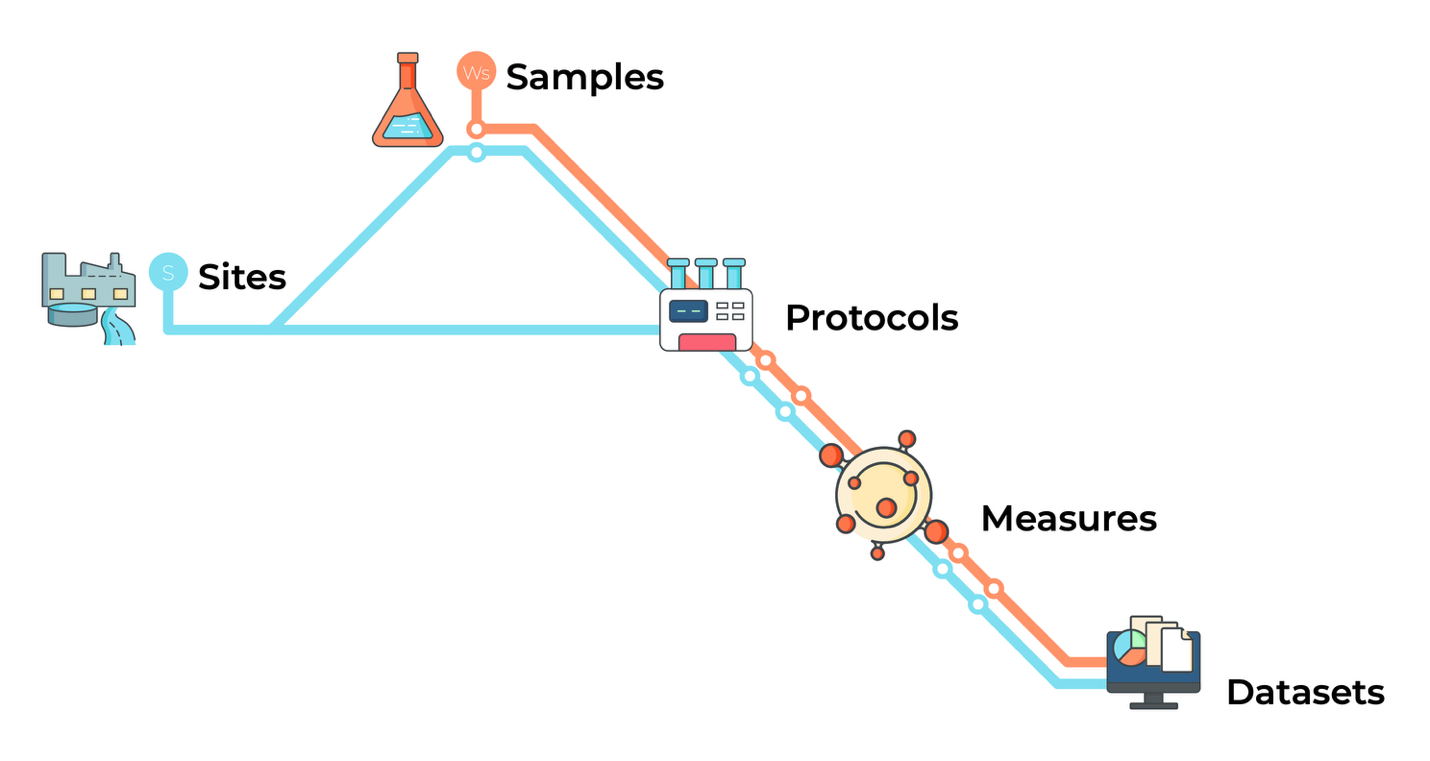}

}

\caption{\label{fig-1}Flow of data in the PHES-ODM from sampling site
through sample collection to measurement. Site information (geographic
location, flow rate) contextualises measurements. Sample information
(material type) is essential for cross-program comparison. Data flows
from site through samples and protocols to a recorded measurement, or
bypasses samples for direct site measurements (e.g., flow rate,
weather).}

\end{figure}

Beyond just site and sample information, the PHES-ODM allows for the
recording of more expansive geographic data, such as polygon
information, which distinguishes it from most other WWS data standards.
In geospatial terms, a polygon is a closed shape specified by a sequence
of unique coordinate pairs where the first and last pair are the same
{[}28{]}. Within the context of WWS these polygons represent defined
areas such as municipal boundaries, wastewater treatment plant catchment
areas, or health administrative regions. By recording these geographical
descriptors and their nested sites, the model ensures interoperability
with Geographic Information Systems (GIS). This hierarchy allows for
data storage across multiple scales:

\begin{itemize}
\tightlist
\item
  Polygons: viral wastewater measures within a broad catchment area.
\item
  Sites: technical data, such as flow rate in a wastewater treatment
  plant.
\item
  Samples: a discrete sample of water or wastewater, collected at a
  point or time period, including how it was collected, transported, and
  stored.
\item
  Measures: include any measurement from a site or sample. The key
  measure is typically a PCR or genomic sequence of an organism.
  However, measures can include contexts such as temperature,
  inhibitors, or chemicals.
\item
  Populations: public health outcomes, such as hospitalization rates
  within a defined area.
\end{itemize}

\begin{figure}

{\centering \includegraphics{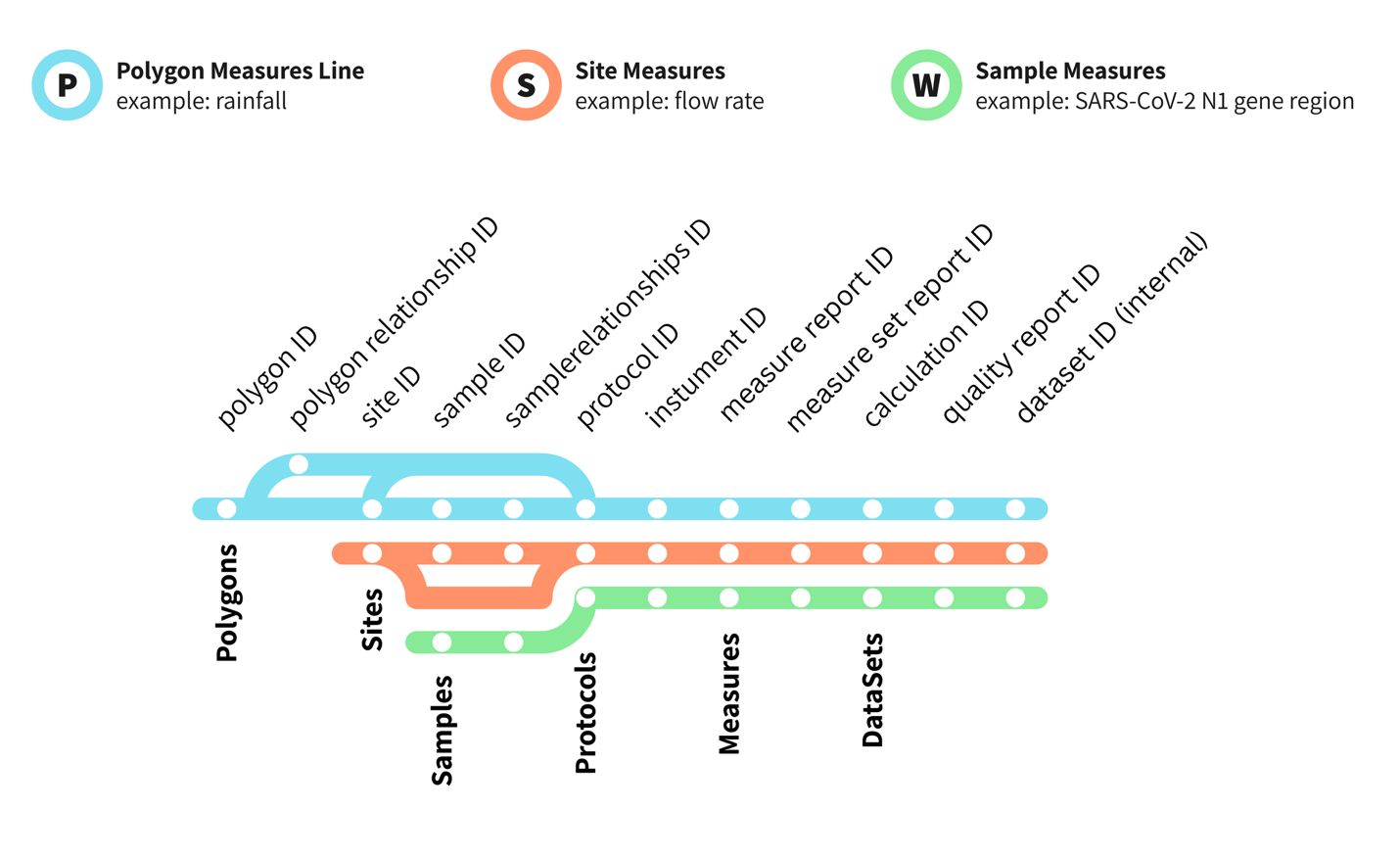}

}

\caption{\label{fig-2}Expanded version of Figure 1, including a data
stream for polygons and the keys used to link data entities within the
model.}

\end{figure}

\begin{itemize}
\tightlist
\item
  The PHES-ODM is implemented as a relational database---a structured
  system in which data are organized into linked tables. The relational
  structure is vital for WWS, where data are inherently complex and
  interrelated. By separately tracking site, sample, population, and
  geographic information, the model provides a comprehensive overview of
  the surveillance landscape while eliminating the onerous data entry
  required for repeated items. In this framework, real-world concepts
  are represented as ``entities'' (e.g., a sample or a site), while
  their specific characteristics are recorded as ``attributes'' (e.g.,
  collection date or building type). In practice, each entity is
  represented by a table, where attributes serve as column headers. The
  relationships between these tables are managed through unique
  identifiers, or keys {[}23-24{]}. Each table in the PHES-ODM has a
  primary key (PK), which uniquely identifies each row in that table.
  When PKs are referenced in another table, to which they are not the
  primary key, they are termed foreign keys (FKs). For example, the
  ``sites'' table records all information about sampling sites, with
  ``siteID'' as its PK. When a sample is collected at a site, ``siteID''
  appears in the ``samples'' table as an FK - linking location data to
  sample records without duplicating site information. This pattern
  repeats: ``sampleID'' serves as an FK in the ``measures'' table,
  chaining sites to samples to measures. For an illustration of this
  structure, see Figure 3.
\end{itemize}

\begin{figure}

{\centering \includegraphics{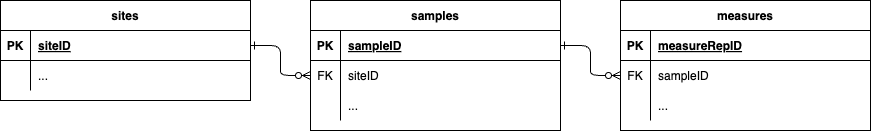}

}

\caption{\label{fig-3}Entity relationship diagram illustrating the
relational structure of the PHES-ODM, showing primary and foreign key
linkages across tables.}

\end{figure}

Separating entities also enables customizability. WWS programs differ in
scale, capacity, and approach; prescribing a single rigid structure
would force local modifications that undermine interoperability. To
address this, the PHES-ODM offers a minimal version containing only
mandatory data and metadata fields, while the full model includes
provisions for any additional items that larger programs may wish to
record. Figure 4 compares the entity relationship diagrams for the
minimal structure (left) and the full version with all optional fields
and reference tables (right).

\begin{figure}

{\centering \includegraphics{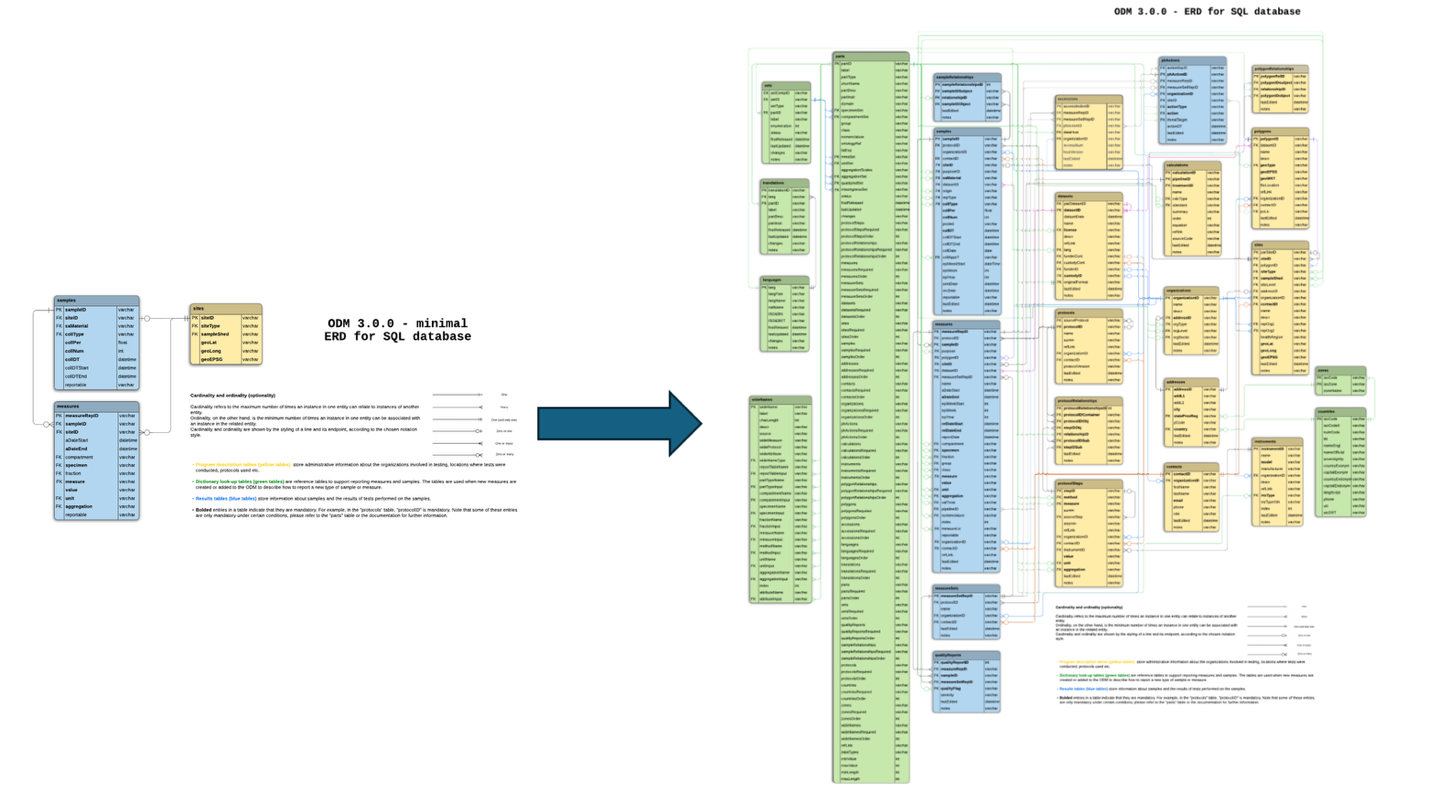}

}

\caption{\label{fig-4}Comparison of entity relationship diagrams for the
minimal PHES-ODM structure (left) and the full model with all optional
fields and reference tables (right).}

\end{figure}

\hypertarget{audience-and-aims-for-wws-data}{%
\subsection{2.2. Audience and aims for WWS
data}\label{audience-and-aims-for-wws-data}}

WWS programs are, by their very nature, complex -- they require
collaboration between engineers, biotechnologists, policymakers, and
public health professionals. The programs require that all these groups
be able to communicate effectively with one another. What is being
communicated between these groups is often domain-specialized knowledge,
making it hard to understand each other's work or share a single
database. Furthermore, different users and analysts may want different
things from the data. By organizing the data around entities, we focus
and streamline data entry by work domain. The model needs to include
features to improve the usability of the data by various actors.

The target audience for the PHES-ODM is thus all these groups:
engineers, biotechnologists, policymakers, and public health
professionals. The model also intends to serve any analysts -whether
primary or secondary -- who work with WWS data.

\hypertarget{wws-data-standardization-and-interoperability}{%
\subsection{2.3. WWS data standardization and
interoperability}\label{wws-data-standardization-and-interoperability}}

The FAIR data principles -- Findable, Accessible, Interoperable, and
Reusable {[}12{]} - are among the most widely cited frameworks for data
standards and data models. Other groups have tried to expand on these
principles, either by adding layers of interoperability {[}29{]}, or by
expanding the principles to encompass other, more human-centric
criteria, such as cognitive interoperability {[}30{]}.

The European Open Science Cloud (EOSC) Interoperability Framework
{[}29{]} specifically posits interoperability as having four layers,
specifically: \textbf{Technical interoperability}, meaning that
Information Technology (IT) systems have completely understood
interfaces and are able to work with other IT systems without
restrictions; \textbf{Semantic interoperability}, which ensures that the
format and meaning of data is preserved and understood throughout
exchanges between parties; \textbf{Organisational interoperability},
meaning organisations align their business processes and
responsibilities to achieve shared goals; and \textbf{Legal
interoperability}, which ensures that organisations can work together
even if operating under different legal frameworks. Vogt's additional
data principles centre largely around ensuring data are
\emph{cognitively interoperable}. That is, data must be understandable
and usable by humans, not only machines {[}30{]}.

Researchers examining secondary use of population data - the situation
facing most WWS analysts who are not the primary data generators -
highlight the importance of robust metadata that makes clear the scope,
limitations, and transformations already applied to the data, alongside
a comprehensive data dictionary that defines variables and supports
validity assessment {[}31{]}. Data standards, and the data dictionaries
that define them, improve interoperability by enabling users to
structure data and metadata in technically and semantically consistent
ways. Yet the development of standards is not without problems. Some are
proprietary, limiting adoption. Even among open standards, overlapping
items are not always interoperable - open standards may be necessary for
health data interoperability, but they are not sufficient on their own
{[}32{]}. Across domains, data standards fall short of their
interoperability goals when they do not actively prioritise
interoperability with one another {[}33{]}. The solution is not a
single, universal standard, but rather several major standards that can
interoperate. Building on that line of reasoning, sound structural
principles make for a good data standard, but so does working with other
standards in the field. The PHES-ODM seeks to apply the interoperability
principles outlined above while also working with other major WWS data
standards to ensure data can flow between them (an overview comparing
the PHES-ODM with other major WWS data standards is provided in Table
1). Interoperability offers tangible benefits---greater data access,
more usable data, improved efficiency, and enhanced collaboration
{[}34{]}. In public health surveillance, these benefits translate into
lives saved.

\hypertarget{structural-solutions-for-data-challenges-and-interoperability}{%
\subsection{2.4. Structural solutions for data challenges and
interoperability}\label{structural-solutions-for-data-challenges-and-interoperability}}

Data standards and dictionaries are designed to ensure that data are
generated, used, and shared without issue, but problems persist due to
structural issues in the data standards, or user error in their
application. Literature on issues encountered in population and public
health data is strong, and the issues persist across research
disciplines {[}31, 35-36{]}. Of particular concern are challenges
around: ambiguous data definitions; a lack of contextual information
(methods applied, standardization of data definitions, location data);
lack of clarity on the temporality of the data; ambiguous data
revisions, or no capacity to perform data revisions; unreported
differences in the level of spatial and temporal resolution across the
data; uncertain data quality; ambiguous ownership, credit, and licensing
of data; and difficulty in databases and data infrastructure around
balancing robustness against ease of use.

The super-issue that underlines most of these problems is one of
ambiguity. With data, and secondary use of data, we do not know what we
do not know. Clearly recording the maximum information is the only
solution for avoiding misuse and misrepresentation of population and
surveillance data. The issue exists throughout the lifecycle of the
data, and if data collectors are unaware of what context is important,
it will not get recorded. The PHES-ODM has thus tried to address these
challenges structurally, by explicitly including provisions to counter
these important ambiguities:

\begin{itemize}
\tightlist
\item
  Data dictionaries and ontology integration to counter ambiguous data
  definitions
\item
  Protocols and calculations data tables to counter a lack of contextual
  information
\item
  Data relevancy periods to clarify data temporality
\item
  ``Last edited'' and ``notes'' fields for tracking data corrections
  transparently
\item
  A ``site level'' and ``specimen'' field for tracking spatial
  resolution of the data
\item
  ``Reportable'' and ``quality flag'' fields for recording measure
  quality issues; a validation library to improve data quality issues
\item
  ``License ``fields connected to datasets and measures to ensure
  responsible and legal use
\item
  Documentation and online community resources to balance ease of use
  against a robust model
\end{itemize}

The other barrier to interoperability through data standards is a lack
of adoption. Many researchers argue for the importance of data standards
and interoperable data, but few implement standards in their work.
Within the GLOWACON technical working group on data, a survey of the
membership indicated the majority thought interoperability and data
standards were important, but almost none used a data standard
themselves. This is a complex problem that is related to a myriad of
factors (lack of awareness, lack of implementation tools; lack of time),
and the PHES-ODM is trying to address them all through:

\begin{itemize}
\tightlist
\item
  Outreach and extensive documentation
\item
  Building tools with built-for-purpose interfaces, including data
  parsers and validators, and
\item
  Ready-to-use templates for common use cases, and video documentation
  to support their uptake.
\end{itemize}

\textbf{Table 1: An overview of the most commonly referenced data models
for wastewater and/or environmental surveillance, and a comparison
between them.} This table is based on Therrien et al's Table 1 {[}14{]}
but is expanded and updated to include additional categories and reflect
the current landscape of environmental public health data models.

\begin{longtable}[]{@{}
  >{\raggedright\arraybackslash}p{(\columnwidth - 14\tabcolsep) * \real{0.1250}}
  >{\raggedright\arraybackslash}p{(\columnwidth - 14\tabcolsep) * \real{0.1250}}
  >{\raggedright\arraybackslash}p{(\columnwidth - 14\tabcolsep) * \real{0.1250}}
  >{\raggedright\arraybackslash}p{(\columnwidth - 14\tabcolsep) * \real{0.1250}}
  >{\raggedright\arraybackslash}p{(\columnwidth - 14\tabcolsep) * \real{0.1250}}
  >{\raggedright\arraybackslash}p{(\columnwidth - 14\tabcolsep) * \real{0.1250}}
  >{\raggedright\arraybackslash}p{(\columnwidth - 14\tabcolsep) * \real{0.1250}}
  >{\raggedright\arraybackslash}p{(\columnwidth - 14\tabcolsep) * \real{0.1250}}@{}}
\toprule\noalign{}
\begin{minipage}[b]{\linewidth}\raggedright
\textbf{Feature}
\end{minipage} & \begin{minipage}[b]{\linewidth}\raggedright
\textbf{PHES-ODM}
\end{minipage} & \begin{minipage}[b]{\linewidth}\raggedright
\textbf{NORMAN SCORE}
\end{minipage} & \begin{minipage}[b]{\linewidth}\raggedright
\textbf{W-SPHERE}
\end{minipage} & \begin{minipage}[b]{\linewidth}\raggedright
\textbf{NWSS}
\end{minipage} & \begin{minipage}[b]{\linewidth}\raggedright
\textbf{PHA4GE}
\end{minipage} & \begin{minipage}[b]{\linewidth}\raggedright
\textbf{AMELAG}
\end{minipage} & \begin{minipage}[b]{\linewidth}\raggedright
\textbf{MIxS}
\end{minipage} \\
\midrule\noalign{}
\endhead
\bottomrule\noalign{}
\endlastfoot
\textbf{Reference} & Manuel et al, 2021 {[}13{]}; Therrien et al, 2024
{[}14{]} & NORMAN Network, 2020 {[}37{]} & Global Water Pathogens
Project, 2020 {[}38{]} & USCDC, n.d. {[}16{]} & Griffiths et al, 2022
{[}39{]}; Paull et al, 2025 {[}40{]} & RKI, 2025 {[}41{]}; RKI \& UBA,
2026 {[}42{]} & Genomic Standards Consortium, n.d. {[}43{]} \\
\textbf{Intended Audience} & WWS practitioners (public health
authorities, Engineers in WWS) & Ecotoxicologists, SARS-CoV-2 template
for WBE practitioners & WWS practitioners & WWS practitioners &
Environmental Genomics & WWS practitioners & Environmental Genomics \\
\textbf{Documentation language} & Narrative documentation: English;
database definitions and data dictionary: English, French, Spanish,
Portuguese & English & English & English & English & German, English &
English \\
\textbf{Public data dictionary of headers and tables} & Yes & No & Yes &
No (outdated version available from USCDC archive) & Yes & Yes & Yes \\
\textbf{Public data dictionary of values} & Yes & No & No & No (outdated
version available from USCDC archive) & Yes & No & Yes \\
\textbf{Database structure type} & Relational database & Flat-file
database & Flat-file database & Flat-file database & Flat-file database
& Flat-file database & Flat-file database \\
\textbf{Public database definition} & Yes & Yes & No & No & Yes & Yes &
No \\
\textbf{Public data conversion tools} & Yes & No & No & No & Yes & No &
No \\
\textbf{Public data validation tools} & Dictionary, software tool,
templates, validation rules schema & Template & Template & None &
Dictionary, software tool, templates & Dictionary, validation rules
schema & Dictionary, software tool \\
\textbf{Public data sharing infrastructure} & Python library, explicit
measure and dataset licensing, external dataset linkages & No &
High-level dashboard & High-level dashboard & Software tool
(DataHarmonizer) & Dashboard, Zenodo publication of data & Many
repositories require adherence to this standard to share \\
\textbf{Public data collection templates} & Yes & Yes & Yes & No & Yes &
No & No \\
\textbf{Governance and development} & Open source & Inter-institutional
& Internal & Internal & Open source & Internal & Open source \\
\textbf{Model license} & CC-BY4 & Not found for the model, but the
template is open access & Not found & Not found & CC-BY4 & CC-BY4 &
CC-BY4 \\
\textbf{Clear channels for user feedback} & GitHub issues, Discourse
discussion board, email maintainers & Email maintainers & Email
maintainers & Email maintainers & GitHub issues, email maintainers &
Email maintainers & GitHub issues, email maintainers \\
\textbf{Rights management} & Element-level (any row, header or
combination) & Dataset level & Dataset level & Dataset level & Dataset
level & Dataset level & Dataset level \\
\textbf{Environmental compartments} & Various & Various, but only
wastewater in the template & Wastewater & Wastewater & Various &
Wastewater & Various \\
\textbf{Pathogen measurement} & Any pathogen in the dictionary
(Multiple+) & Yes, but only SARS-CoV-2 in template & SARS-CoV-2-specific
& Multiple & Multiple & Multiple & Multiple \\
\textbf{Detailed protocol recording and linkage} & Yes & No & No & No &
No & No & No \\
\textbf{Detailed sample relationship records} & Yes & No & No & No & No
& No & No \\
\textbf{Measurement methods} & Yes & Yes, but only PCR and
sequencing-specific in the template & PCR and sequencing-specific & PCR
and sequencing-specific & Sequencing-specific & Not found &
Sequencing-specific \\
\textbf{In-sample measurements} & Any measure in the dictionary & Water
quality, but only PCR in the template & PCR and sequencing & PCR and
sequencing, pH, Conductivity, TSS & PCR and sequencing, water quality &
PCR and sequencing, pH, temperature & PCR and sequencing, water
quality \\
\textbf{Collection site information} & Yes & Yes & Yes & Yes & Yes & Yes
& WWTP infrastructure details \\
\textbf{On-site measurements} & Any measure in the dictionary
(expandable) & Flow, Weather, COD, TSS, NH4+-N, Water temperature & Flow
& Flow, water temperature & Flow, Weather, COD, TSS, NH4+-N, Water
temperature, conductivity, pH, contamination & Flow, pH, temperature &
COD, TSS, NH4+-N, phosphate, salinity, \\
\textbf{Population count} & Served by site, or within a geographic
region (polygon) & Served by site & Served by site & Served by site &
Served by site & Served by site & No \\
\textbf{Sewer network information} & Possible to record details as
measures in the dictionary & No & No & Average wastewater travel time,
industrial input, stormwater input & Upstream activity and treatment &
No & Industrial input, reactor type, sludge retention time \\
\textbf{Sample and sampling method} & Yes & Yes & Yes & Yes & Yes & No &
No \\
\textbf{Used by a national/ supranational WWS program} & Yes & No & No &
Yes & No & Yes & No \\
\textbf{Records provenance and transformation steps} & Yes & No & No &
No & Yes (accession IDs for reference sequences, sequences; libraries \&
processing software) & Yes (reports viral load, flow-standardized viral
load, and predicted viral load) & Yes (libraries \& processing
software) \\
\textbf{Genomic repository linkages} & Yes & No & No & No & Yes & No &
Yes \\
\textbf{Population health data} & Any measure in the dictionary
(aggregate health data -- population level) & SARS-CoV-2 prevalence & No
& No & No & No & No \\
\textbf{Ontology integration} & Limited & No & No & No & Yes & No &
Yes \\
\textbf{Interoperable with at least one other major dictionary using
public tools} & Yes (PHA4GE, NWSS) & No & No & Yes (PHES-ODM; managed by
PHES-ODM) & Yes (PHES-ODM; managed by PHES-ODM) & No & No \\
\end{longtable}

\hypertarget{results-discussion}{%
\section{3. Results \& Discussion}\label{results-discussion}}

\hypertarget{addressing-the-audience-public-health-surveillance}{%
\subsection{3.1. Addressing the audience: public health
surveillance}\label{addressing-the-audience-public-health-surveillance}}

With WWS, we are surveilling the environment for evidence of infection
and threats to human health. As a disease surveillance system, it exists
under the umbrella of public health. The data generation, however, comes
from sites and sampling that exist outside of the typical
(i.e.~clinical) setting for public health surveillance. Most of the
research, testing, and the programs, at least in Ontario, are or were
run by engineering and environmental science laboratories {[}44{]}. This
ensures relevant contextual data around sites, samples, and wastewater
processing are collected, and that the best information about the
presence of pathogens and other threats to human health were being
extracted from the complex sample matrix that is wastewater. This
context is unfamiliar to public health practitioners, however, and
translation across each discipline's dialect is often required.

Version 1 of the PHES-ODM was developed in haste with partners and
health authorities to create an open data standard at the height of the
COVID-19 pandemic response. The infrastructure to facilitate
communication between engineers and public health teams at that point
was including provisions to store case, hospitalization, and death data
in the same dataset as WWS data. As the program developed, however,
needs changed. As public health departments started predictive modelling
projects, alerting them to which measures to use, or which not to use,
became very important. In version 2 of the PHES-ODM, more robust quality
flagging was added for measures and samples. More importantly for public
health departments and analysts, however, was the new ``severity''
attribute to alert them to how important a quality issue was. A binary
(``TRUE'' or ``FALSE'') ``reportable'' field was also added to both the
``measures'' and ``samples'' tables so analysts can tell immediately
whether the data is dependable. The data dictionary (the ``parts''
table) is also available to provide additional guidance on the meaning
of terms.

\hypertarget{the-public-health-actions-table}{%
\subsubsection{3.1.2. The public health actions
table}\label{the-public-health-actions-table}}

Building on previous work, version 3 of the PHES-ODM maintained the
original features to support public health departments in understanding
WWS data or expanded them. The public health aspect of the data has also
been made even more explicit by adding the optional ``Public Health
Actions'' table to the model. The purpose of the ``Public Health
Actions'' table is to link health-related policy actions to the measures
that inform and/or result from them.

The table and the values for its enumerated fields can be found in
\textbf{Figure 5a}. ``phActionID'' serves as the primary key for the
table and is a unique identifier for each row or action taken. If
multiple actions are being taken at the same time, a second
``phActionID'' can be generated as an umbrella action ID, and the
``actionType'' and ``action'' fields left blank. The actual entries for
each of the related actions will then use the ``phActionID'' of this
partially blank row as the action group ID (``actionGrpID'') (see
\textbf{Figure 5c} for an example). The measures, or group of measures,
related to the public health action are linked to it by their IDs in the
``measureRepID'' or the ``measureSetRepID'' field, where they act as
FKs. Currently, the causal relationship between measures and actions is
left ambiguous; the fact that there is a relationship is specified, but
it is not made clear if the action is a result of the linked measure(s),
or whether the measure(s) are a result of the action. The organization
responsible for the public health action is linked using the
``organizationID'' as an FK from the ``organizations'' table. The site,
if relevant (particularly for a hospital or clinical setting, for
example) can be linked, using the ``siteID'' as an FK from the ``sites''
table. The ``actionType'' field serves as a high-level general organizer
for the type of action being undertaken, while the ``action'' field adds
more detail. ``threatTarget'' is the pathogen being targeted by the
action, and it is populated by ``measurement'' part types from the parts
table (i.e.~items like ``sarsCov2'', ``fluA'', etc.). In instances with
multiple targets, each pathogen will need an individual row, and they
can be grouped using ``actionGrpID''. The date and time of the public
health action (``actionDT'') is recorded in this table, along with a
relevance start and end date (``relDateStart'' and ``relDateEnd'') to
mark the projected period of activity for a given action. Finally, the
``lastEdited'' field and the ``notes'' field record when (if ever) a row
in the table was updated, and any additional details about that action,
respectively.

\textbf{Figure 5b} and \textbf{Figure 5c} walks through the following
examples for using the table: a public health department (phd) for City
A issues a masking recommendation (``maskRec'') as an infection control
measure (``contMeasImp'') for influenza A virus (``fluA''); and that
same public health department in that same city declares the start of an
outbreak (``outbStart'') as part of an outbreak alert (``outb'') for
respiratory syncytial virus B (``rsvB''), and increasing surveillance of
the pathogen (``incSurv'') as part of a surveillance alert
(``survAlert'').

\begin{figure}

{\centering \includegraphics{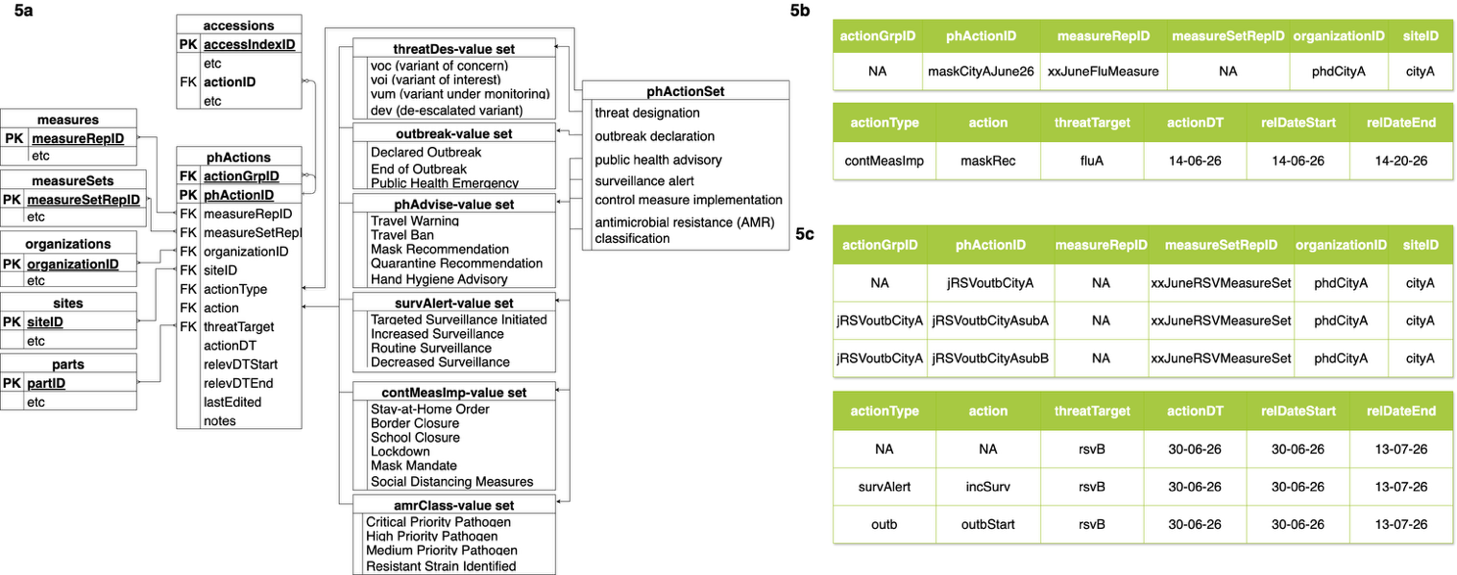}

}

\caption{\label{fig-5}Entity relationship diagram of the phActions table
and its integration into the PHES-ODM (Figure 5a), with example data
entries for a single public health action (Figure 5b) and a grouped
multi-pronged action (Figure 5c).}

\end{figure}

\hypertarget{addressing-the-audience-data-analysts}{%
\subsection{3.2. Addressing the audience: data
analysts}\label{addressing-the-audience-data-analysts}}

Across the data lifecycle, users need data organized differently during
analysis than when stored in a database. Moving between these two
formats, however, may introduce errors and cause friction. To support
reproducible data transformation and avoid errors, the PHES-ODM supports
a ``wide'' data format. ``Wide'' and ``long'' formatted data are more
general descriptors than fixed categories, with all tabular data
existing somewhere between the two. For our purposes here, we will
define ``long'' data as a format where one row represents a single
entity, while ``wide'' data uses columns to store information about
multiple measures or entities per row, usually for a shared entity like
a date or a site. An example comparison is shown in \textbf{Figure 6a},
where in the ``long'' format the measures for a date have their own
rows, and in the ``wide'' format each date is a row, with additional
columns for each measure. The standard PHES-ODM format is a ``long''
data format, which is ideal for storage and database scalability, and
ensures that the data are very machine-actionable. For larger analyses
and for faster data entry templates, however, a ``wide'' format is often
preferred. Though, while convenient for quick data entry, the
compactness of wide tables leaves little room for metadata beyond the
column header.

\begin{figure}

{\centering \includegraphics{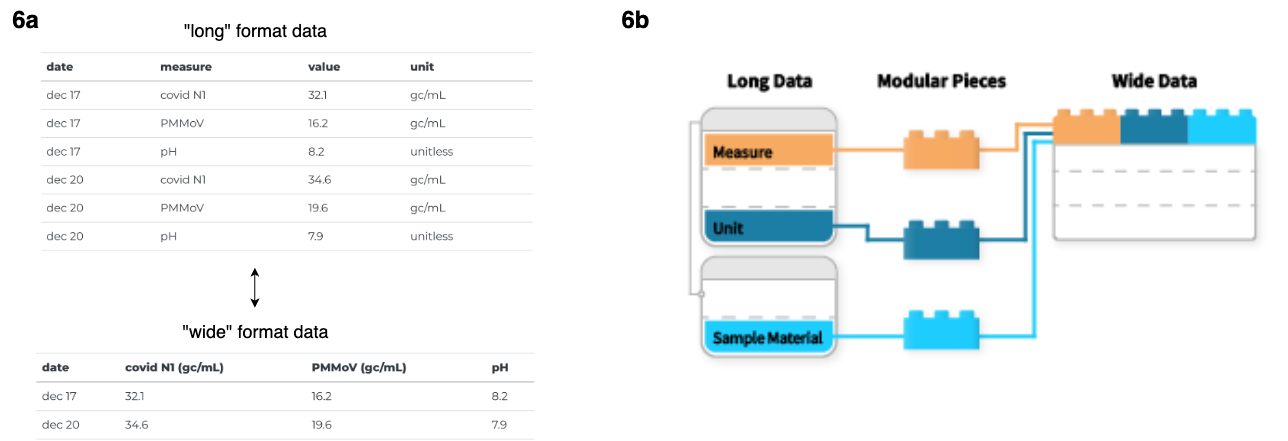}

}

\caption{\label{fig-6}Example of the same measurement data in ``long''
vs ``wide'' formatting (Figure 6a), and illustration of the modularity
of the relational PHES-ODM data structure showing how attributes from
different tables can be combined into customised templates (Figure 6b).}

\end{figure}

The generation of wide names and their use in templates showcases the
modularity of the PHES-ODM model. The model is very large in order to be
robust and to accommodate as much data for as wide an array of potential
user as possible. The robustness is meant to support compliance to the
model and avoid having users modify the model locally to respond to a
need (creating problems for interoperability). Most fields in the model
are, however, optional. The full model should be considered like a menu
at the restaurant, or a bucket full of Lego bricks; all these items are
available to you in their standard format, and it is up to you (and the
needs of your program) to determine what you select, adopt, and use.
This applies to both the long and wide version of the model. An
illustration of this kind of modularity is shown in \textbf{Figure 6b}.

\hypertarget{addressing-standardisation-challenges-data-mapping-and-interoperability}{%
\subsection{3.3. Addressing standardisation challenges: data mapping and
interoperability}\label{addressing-standardisation-challenges-data-mapping-and-interoperability}}

\begin{figure}

{\centering \includegraphics[width=0.45\textwidth,height=\textheight]{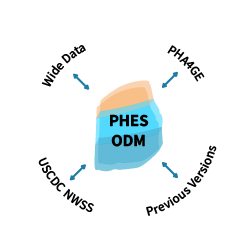}

}

\caption{\label{fig-7}The PHES-ODM as a Rosetta Stone between wastewater
surveillance data standards, with mapping tools enabling ingestion of
PHA4GE, USCDC NWSS, and earlier PHES-ODM formats.}

\end{figure}

As mentioned above, achieving a single, unified data standard for WWS is
an unrealistic expectation. Developing open standards, like the
PHES-ODM, is part of the solution, but is insufficient. There is a
proliferation of non-interoperable systems within and without the field
of WWS. The developers of standards and models need to prioritize
interoperability, and to work to build tools to interface with other
systems and standards {[}33{]}. To this end, the PHES-ODM aims to serve
as a Rosetta Stone (\textbf{Figure 7}) between other WWS data standards
and models. Currently, data in the PHA4GE data format {[}39-40{]}, the
USCDC NWSS data format {[}16{]}, as well as previous versions of the
PHES-ODM can all be mapped into version 3 of our model using tools
developed by our team {[}46{]}.

The easiest part of interoperability are the basic objects, such as
categorical inputs, unique identifiers, and date fields --- perhaps in
part because these objects tend to be non-ambiguous. The main struggle
here is to ensure data being brought into the PHES-ODM format has a
destination field that matches. Ensuring that the PHES-ODM has overlap
with other WWS data models has, for example, led to a proliferation of
date-type fields. Because different labs and standards record date
information in different ways, version 3 of the PHES-ODM now includes
other options for how to record dates and times. These alternatives
include recording the epidemiological week {[}47{]}, including the start
date and year of that epidemiological week; and recording the date with
a categorical generalization of when the sample was collected
(i.e.~morning, afternoon, evening, night). Beyond these easier targets,
metadata can be difficult to match across standards due to different
scales of recording and measurement, structural differences, or
imperfect semantic overlap. This is not a novel problem, and natural
language translation has worked around imperfect translation since time
immemorial. The problem arises when imperfect translation is left
ambiguous or unexplained. To help explain possible irregularities
generated by mapped and transformed data, version 3 of the PHES-ODM has
a field in the ``datasets'' table for recording the ``originalFormat''
of the data, also supporting greater data transparency.

\begin{figure}

{\centering \includegraphics{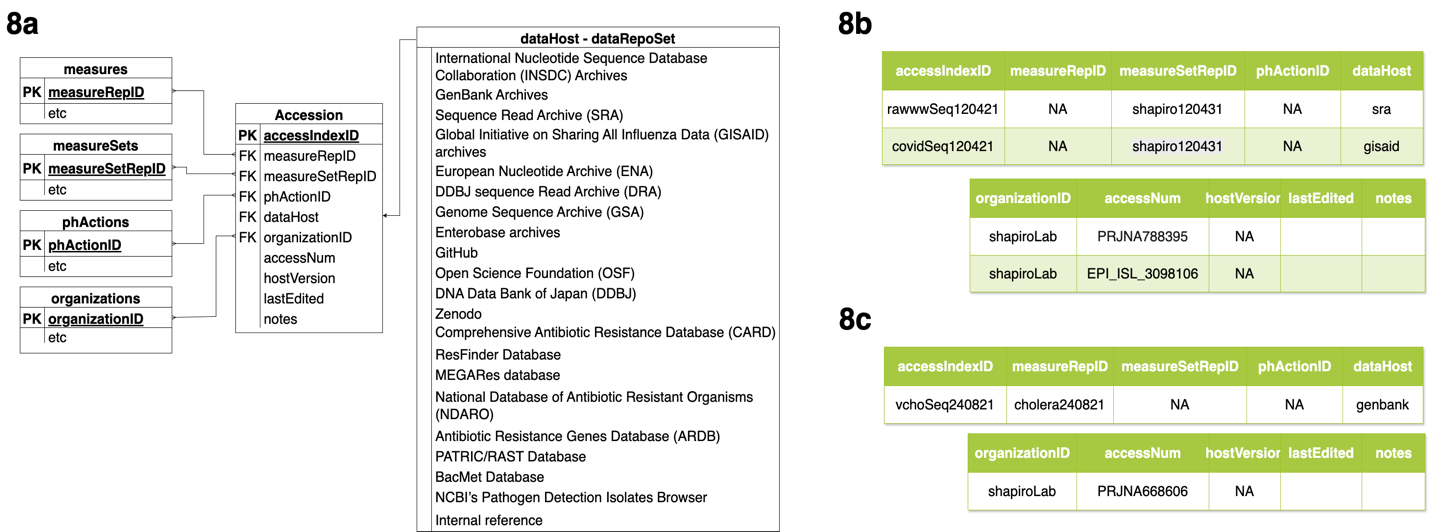}

}

\caption{\label{fig-8}Entity relationship diagram of the accessions
table (Figure 8a), with example data entries illustrating linkages to
external sequencing repositories and public health records (Figures
8b--8c).}

\end{figure}

Linking out to the dataset in its original format is also possible
thanks to version 3's new optional ``accessions'' table. This table is
designed to link external data, whether that be original data sources;
large data that cannot be stored well outside its context
(i.e.~sequencing data, GIS data); or public health data and dashboards.
An entity relationship diagram of the table can be found in
\textbf{Figure 8a}. The ``accessionIndexID'' is the PK for the table and
is a unique identifier for each row. The ``measureRepID'' or
``measureSetRepID'' link as FKs to a single measure or sets of measures
related to the external data being referenced. For example, if a measure
is reporting the proportion of a given variant found in a sample of
wastewater, that measure can be linked to the accessions to point out to
the full sequencing data from that finding. The ``phActionID'' works
like the measure identifiers, but links public health actions to
external data on the details of that action. ``dataHost'' is a
categorical variable for reporting to what repository the accession is
pointing. For internal databases linking across different departments or
sectors, an ``internal reference'' category is also available. The
``organizationID'' field links to the organization that is associated
with the external data. For example, users may point out to a reference
sequence generated by another group that was used to confirm their
findings. The ``accessNum'' field is a free text field for reporting the
accession number or ID for an entry in the ``dataHost'' repository.
``accessNum'' can also take a web address as an input. To report on
different repository versions, where applicable, ``hostVersion'' can be
used to record that data. Lastly, as with all PHES-ODM tables, the
optional ``lastEdited'' field and ``notes'' fields record when data were
last updated, and any other details, respectively.

\textbf{Figure 8b} and \textbf{Figure 8c} provides a fictional example
of how accessions data might be recorded using examples constructed from
references in two papers from the Shapiro lab {[}48-49{]}.
\textbf{Figure 8b} shows two accessions linked to the same set of
measures as an example, with one pulling the SRA database Bioproject
accession for the raw wastewater sequencing data (\textbf{Figure 8b},
row 1), and the viral genomes from clinical samples in GISAID
(\textbf{Figure 8b}, row 2). \textbf{Figure 8c} shows linkages to a
single measure, for metagenomic sequence data that are available in
GenBank.

\hypertarget{implementing-structural-solutions-expanding-metadata-and-their-context}{%
\subsection{3.4. Implementing structural solutions: expanding metadata
and their
context}\label{implementing-structural-solutions-expanding-metadata-and-their-context}}

The bulk of the PHES-ODM is metadata. The actual measures take up very
few fields, while provisions for contextual information are numerous.
This is in part due to the complexity of environmental surveillance,
which requires extensive metadata to make the information found in
measures useable. Ensuring that the data generated by these programs are
useable is also a core responsibility of WWS programs, as a part of
managing data ethically and upholding principles of data justice. This
does not mean all data should be publicly available -- while no exact
number has been verified, and community shedding dynamics likely differ
widely, some programs are already withholding data about smaller
sewersheds (\textless3000 people) on the grounds that they may not
benefit from a large enough pool for the data to be anonymized
{[}50-51{]}. This means that some WWS data is sensitive personal health
data. For sites in large urban centres, however, this is not an issue.
While Brown et al state that the three things that matter in science are
the data, the methods for data collection, and the logic that connects
the former two to their conclusions {[}52{]}, the greater the context
and metadata provided, the greater the logic to ground one's
conclusions.

Metadata fields added in version 3 of the PHES-ODM were primarily added
at the request of users. Others were added to provide structural
solutions to the common data problems mentioned in methods section 2.4,
Structural solutions for data challenges \& interoperability. The
impetus for structural solutions is that discussion of data problems
does not necessarily provide solutions; acknowledging a problem is not
enough. Some of the structural solutions have been a part of the model
across several versions, but their specific use will be discussed here.
By building in specific metadata fields to eliminate these problems, the
aim is to support users in generating better data.

\hypertarget{implementing-structural-solutions-data-definitions}{%
\subsubsection{3.4.1 Implementing structural solutions: data
definitions}\label{implementing-structural-solutions-data-definitions}}

Issues around ambiguous data definitions are alleviated by the data
dictionary, which has been a part of all versions of the PHES-ODM. By
providing explicit definitions and instructions for every field and
value within the model, ambiguity is resolved. The use of controlled
vocabularies, like ontologies, are also a great boon to ensuring shared
definitions. For version 3 of the PHES-ODM, the ontology integration was
expanded and will increase with successive releases. Some data
definitions have expanded or shifted across version releases of the
PHES-ODM. Using ontology integration helps to limit issues caused by
these updates, but the model also includes changelogs between versions.

\hypertarget{implementing-structural-solutions-defining-temporality-of-data}{%
\subsubsection{3.4.2 Implementing structural solutions: defining
temporality of
data}\label{implementing-structural-solutions-defining-temporality-of-data}}

Temporality issues are another challenge addressed in the PHES-ODM. A
feature of the model that has existed across multiple releases is the
``lastEdited'' field in all tables. This field allows users and data
managers to update data entries to correct errors as needed, while
making the correction process transparent. A similar issue for
temporality is recording a context window for data. For example, the
measures of population size are generated from census data which are
used for five years, or until the next census results are released.
Editing the population counts for every census may change the
denominator for past calculations and lead to erroneous conclusions.
Providing a context window using relevance start and end dates
(``relDateStart'', ``relDateEnd'') to certain tables allows users and
data generators to define the periods for which certain data apply.

\hypertarget{implementing-structural-solutions-contextualizing-data-quality}{%
\subsubsection{3.4.3 Implementing structural solutions: contextualizing
data
quality}\label{implementing-structural-solutions-contextualizing-data-quality}}

Data quality issues are a perennial problem when working with data. As
discussed above, some data quality concerns can be managed by using
quality flags and the ``reportable'' field. This records quality issues
at the level of the sample, the measure, or the methodology, along with
an indicator as to their severity. It also provides, via the
``reportable'' flag, a quick binary flag for whether a measure or sample
should be included in calculations or reports. Data entry or validity
errors are, however, not possible to address or catch in this way. To
resolve these issues, however, our group developed a validation tool
using Python {[}53-54{]}. This tool has detailed documentation and
returns a report on data validity issues.

\hypertarget{implementing-structural-solutions-ownership-and-licensing}{%
\subsubsection{3.4.4 Implementing structural solutions: ownership and
licensing}\label{implementing-structural-solutions-ownership-and-licensing}}

Data ownership and legal interoperability is a major concern in research
- particularly around concerns of ``getting scooped'' on data
publications - and consideration of proper attributions and data
provenance have been present since the inception of the model. Version 2
of the model included the ``datasets'' table to provide information on
who generated the data. This table has numerous linkages to properly
connect data to their sources. Version 3 builds on this by adding a
``license'' field to record the licensing of a dataset. A measure
license (``measureLic'') field was also added to the measures table to
record licensing at the level of individual measures if necessary. The
suite of tools designed to support the PHES-ODM also includes a sharing
tool {[}55{]} which uses an allow-list approach to automatically filter
data for sharing.

\hypertarget{implementing-structural-solutions-data-treatments-and-the-calculations-table}{%
\subsubsection{3.4.5 Implementing structural solutions: data treatments
and the calculations
table}\label{implementing-structural-solutions-data-treatments-and-the-calculations-table}}

The most complicated addition to the model in version 2 was the three
protocols tables (``protocols'', ``protocolRelationships'', and
``protocolSteps''). While initially complex, these tables can be adapted
to be as simple as users require. Laboratory and sampling protocols are,
however, complicated, and representing complex processes recorded in
natural language as linked rows of data required some creative thinking.
The inclusion of methods information is critical to understanding the
data and provides invaluable context for interpreting and aggregating
measurements. Moving beyond laboratory protocols, version 3 adds the
optional ``calculations'' table to the model. This addition makes
transparent what transformations and calculations have already been
applied to the data, and what mathematical and analytical methods were
applied to a measure or aggregation. Because the PHES-ODM allows users
to record both raw measurements, as well as aggregations and estimates,
being able to robustly and transparently report how these calculations
were performed is crucial context. An entity relationship diagram of the
``calculations'' table can be found in \textbf{Figure 9a}.

Within the calculations table, the ``calculationID'' field is the PK for
the table. It is generated in practice as a composite key, made by
concatenating the ``pipelineID'' and ``treatmentID'' fields. The
``pipelineID'' is the identifier for a data pipeline, or a series of
calculations/data treatments. ``pipelineID'' is also the field that will
link to the ``measures'' table and be referenced there. This ID is the
shorthand for the data pipeline used. ``treatmentID'' is identifier for
each data treatment or calculation within the ``calculations'' table. A
series of treatments make up a pipeline, or a pipeline can be a single
treatment. The fields ``name'' and ``summary'' are optional free text
fields to record a name or to summarize data treatment. If used, the
``summary'' field should explain terms used in the ``equation'' field.
The ``calcType'' field is a categorical variable used to explain the
nature of a data treatment. The valid enumeration values are
``normalization'', ``standardization'', ``smoothing'', or
``predictiveModels''. For the ``standard'' field, users categorically
record the standard to which something is standardized (i.e.~Pepper Mild
Mottle Virus (PMMoV), wastewater flowrate, etc.) or smoothed
(i.e.~Bayesian smoothing, 7-days, time, etc.). The field is populated by
measures and categories available in the larger model. When structuring
data treatments in a pipeline, the ``order'' field uses an integer to
structure the flow of treatments within the pipeline. The ``equation''
field optionally specifies the equation used in the data treatment. The
reference link (``refLink'') field provides a link to details on the
data treatment. ``sourceCode'' records the source code for the data
treatment. This field is more applicable for algorithms and complex
modeling. Users can record the full code as text or enter a URL to where
the code is stored (a different URL than the ``refLink'' field).
Finally, the optional ``lastEdited'' field and ``notes'' fields record
when data were last updated, and other details.

The ``calculations'' table works in concert with the new ``dataTreat''
field in the ``measures'' table. This new field takes enumeration values
and is used as a quick flag to describe the nature of a measure. This
helps avoid issues of analysts accidentally using predicted or estimated
measures as raw or aggregated data.

\begin{figure}

{\centering \includegraphics{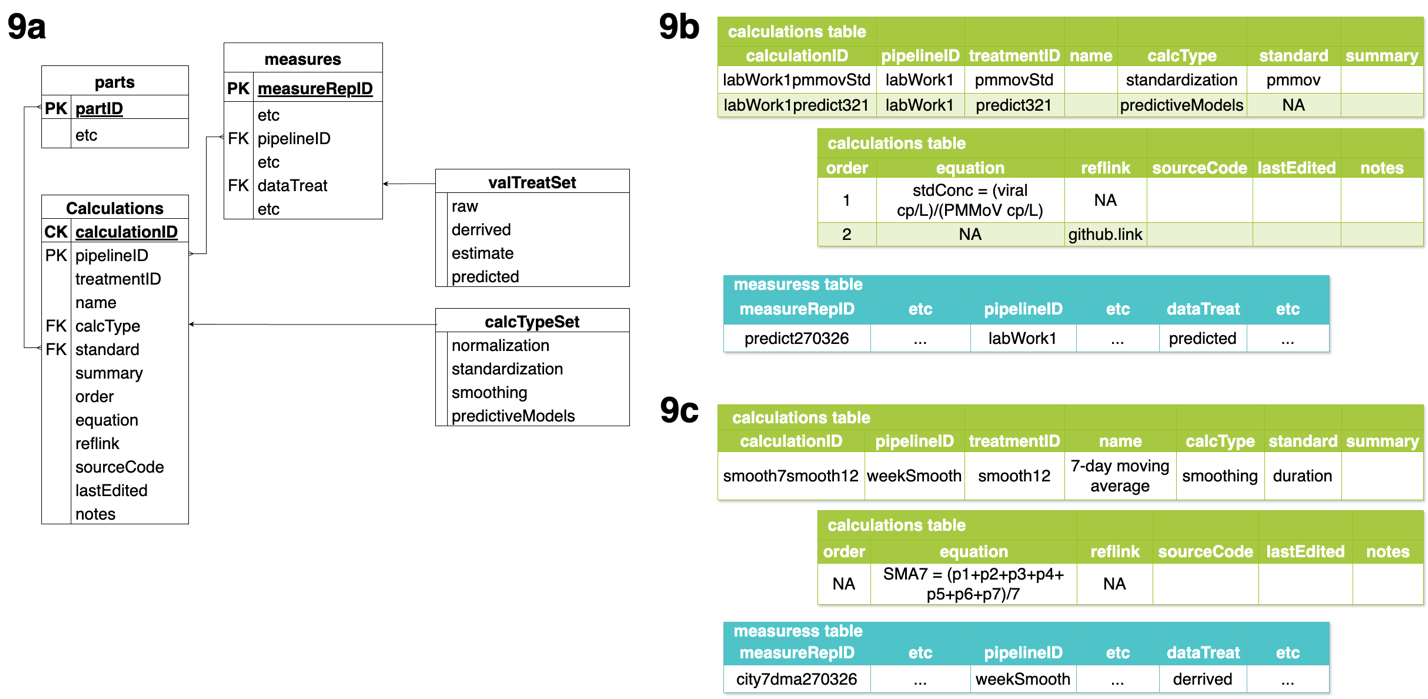}

}

\caption{\label{fig-9}Entity relationship diagram of the calculations
table with example data entries illustrating multi-step and single-step
data treatment pipelines (Figures 9a--9c).}

\end{figure}

As an example, a user recording an estimated measure generated by a
predictive algorithm using WWS data standardized to PMMoV in the
``measures'' table would record the measure normally. The ``dataTreat''
field in the ``measures'' table for that measure would take the value
``predicted''. In the ``calculations'' table, the standardization of the
WWS data to PMMoV would be recorded as one data treatment, with the
``calcType'' field recording ``standardization'' and the ``standard''
field recording ``pmmov''. The ``order'' field in this row would record
``1'', as it is the first step in the pipeline. The predictive algorithm
would be recorded as a second treatment, with the ``calcType'' of
``predictiveModels'', and the ``standard'' left blank or ``NA''. The
``order'' in this row would be ``2'' as it is the second step in the
pipeline. Both rows would share a ``pipelineID''. An example of this
data entry can be found in \textbf{Figure 9b}. Another example, this
time of a single-treatment pipelines, is found in \textbf{Figure 9c},
where a 7-day moving average calculation is done to smooth the WWS data.
In the ``measures'' table, the associated measure would record
``derived'' in the ``dataTreat'' field. In the ``calculations'' table,
the ``calcType'' is ``smoothing'' and the ``standard'' is ``duration''
(time). The ``order'' field is left blank, or with ``NA''.

\hypertarget{implementing-structural-solutions-site-level-and-recording-spatial-resolution}{%
\subsubsection{3.4.6 Implementing structural solutions: site level and
recording spatial
resolution}\label{implementing-structural-solutions-site-level-and-recording-spatial-resolution}}

The issue of unreported differences in the spatiotemporal resolution of
a measure, particularly in aggregated data sets, is an important one.
The new ``calculations'' table helps to some extent by providing details
on standardization or aggregation calculations. It is still, however,
important to know what geography a sample or measure is intended to
reflect. To address this, the PHES-ODM has always stored polygon
information, capturing the exact geography or catchment boundaries for a
site. In situations where polygon information is not available, or where
it is unclear whether the area represented is a city or a region,
problems persist. For example, let us say that wastewater treatment
plant X (WWTP X) is a WWTP that services REGION X: a region in which
there are three municipalities. There is also WWTP Z, which is the
wastewater treatment plant that services REGION Z: a region in which
there are four municipalities. The way that the wastewater
infrastructure is built, it is possible to sample WWTP X such that you
can have measures that reflect each of the municipalities individually,
or the whole region. Conversely, WWTP Z can only be sampled and measured
in a way that reflects the whole region, not individual municipalities
(\textbf{Figure 10a}). When it comes to analysis and aggregation, it is
useful to users to be able to differentiate between the geographic level
that is being sampled (whether that's a region, or something smaller) so
that comparisons can be more meaningful. For example, comparing the
measures from Town X2 and Region Z as though they are the same type of
geography would not be appropriate. While comparison at this level is
allowed, making explicit that there is a spatial resolution difference
is essential. This means ensuring future data users know that it is not
possible to get any granular data from Region Z at the municipal level.

To address this, ``siteLevel'' is a new field added to the ``sites''
table to eliminate this kind of ambiguity. The valid categorical values
for ``siteLevel'' are shown in \textbf{Figure 10b} and are: country
level aggregation, the measures at this site reflect an entire country;
province level aggregation, the measures at this site reflect an entire
province or state; A regions, the measures at this site reflect an
entire greater metropolitan area (GMA) made up of several smaller
municipalities; B regions, the measures at this site reflect multiple
municipalities that are a part of a shared GMA, but do not reflect a GMA
in its entirety; municipality level, the measures from this site reflect
a single municipality; neighbourhood level, the measures from this site
reflect a single neighbourhood; and First Nations level, the measures
from this site reflect a single First Nation. An example data entry for
the example in \textbf{Figure 10a} can be found \textbf{in Figure 10c}.

\begin{figure}

{\centering \includegraphics{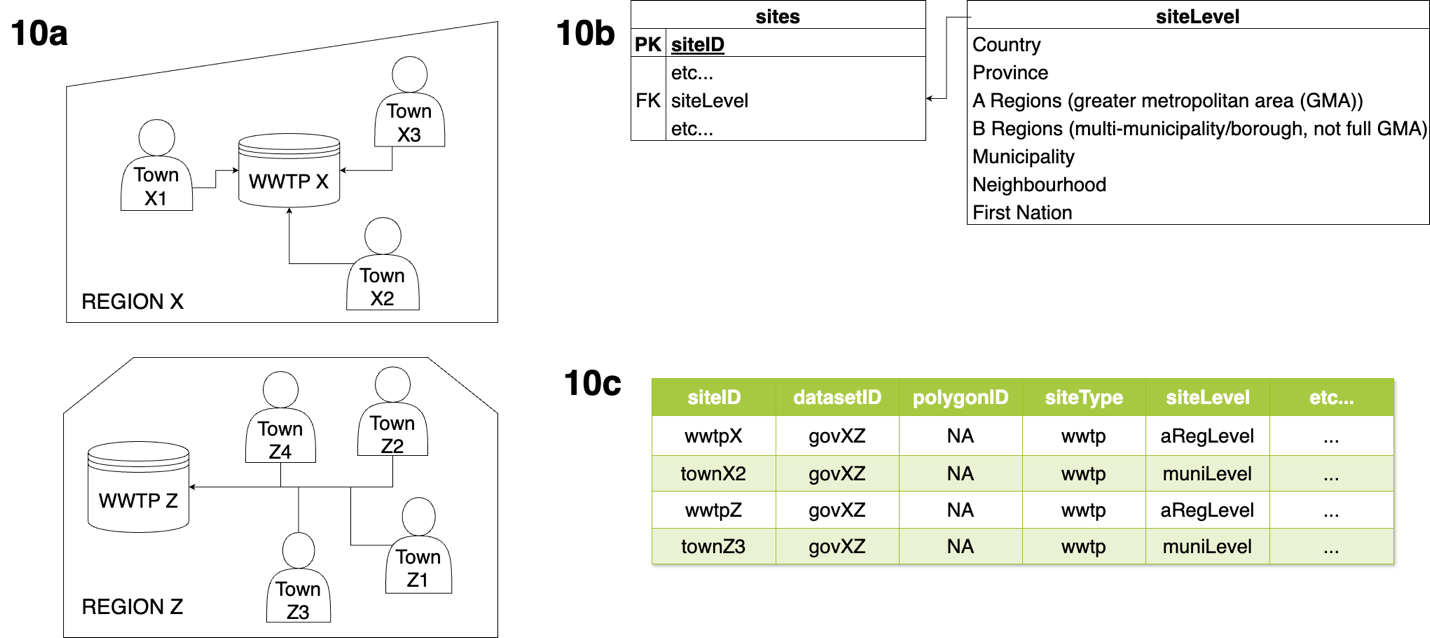}

}

\caption{\label{fig-10}Site-level categorical classification and spatial
resolution examples illustrating the new siteLevel field (Figures
10a--10c).}

\end{figure}

\hypertarget{implementing-structural-solutions-robustness-vs.-ease-of-use}{%
\subsubsection{3.4.7 Implementing structural solutions: robustness
vs.~ease of
use}\label{implementing-structural-solutions-robustness-vs.-ease-of-use}}

One final issue with data models is balancing ease of use with
robustness. This is a challenge for all data standards, and one that we
have tried to balance since the beginning. The first version of the
PHES-ODM was very straight forward and easy to use but was very limited
in what it could report. Version 1 could still work today for very basic
reporting of SARS-CoV-2 detection. As the field of WWS matured, however,
additional targets needed to be added, along with different measures of
the different targets, additional metadata, and more context. With
version 2 and now version 3 of the model, we have striven to be as
robust a model as possible and included everything that was asked of us.
With this, however, the complexity has grown. Today when interacting
with new users it is not uncommon to hear that the model is somewhat
intimidating, and they worry about the time needed to invest to
understand and adopt the model. To respond to this issue, however, we
have made robust documentation website {[}56{]} and video tutorials
{[}57{]} available so that users can feel more comfortable in the model
right away. We also have a message board hosted on Discourse {[}58{]}
where users can ask questions, and communications about new developments
to the model are openly discussed. New issues for new parts can be
submitted on GitHub {[}59{]}, and we are starting to explore
AI-supported tools to empower users to get started with as low of a
barrier as possible.

The PHES-ODM has structural provisions to address issues around data
definitions, and around balancing robustness against ease of use. These
are addressed through the larger support structure around the model,
rather than necessarily the structure of the model itself. This larger
support, documentation, and education ecosystem around the PHES-ODM also
helps to address issues around understanding the data and its primary
use, and around understanding the classification and coding systems of
the values used in the model structure.

\hypertarget{implementing-structural-solutions-expanding-data-relationships}{%
\subsubsection{3.4.8 Implementing structural solutions: expanding data
relationships}\label{implementing-structural-solutions-expanding-data-relationships}}

As a relational database structure, the PHES-ODM works to represent
real-world relationships in the data. In version 2 of the data model, in
addition to the relational linkages between tables, additional tables
were added to record more complex relationships, and many ``parent''
fields were added. Therrien et al cover the structure of the
relationships tables very well {[}14{]}. The parent fields in version 2
(namely Parent Dataset ID, Parent Site ID, and Step Provenance ID)
served a similar purpose to the relationships tables, but on a much
simpler level. They allowed sites or datasets to be contained within
larger sites or datasets, making geography and data ownership levels
more explicit. The Step Prevenance ID also helped to make a type of
citation possible by linking related or adapted protocol steps.

\begin{figure}

{\centering \includegraphics{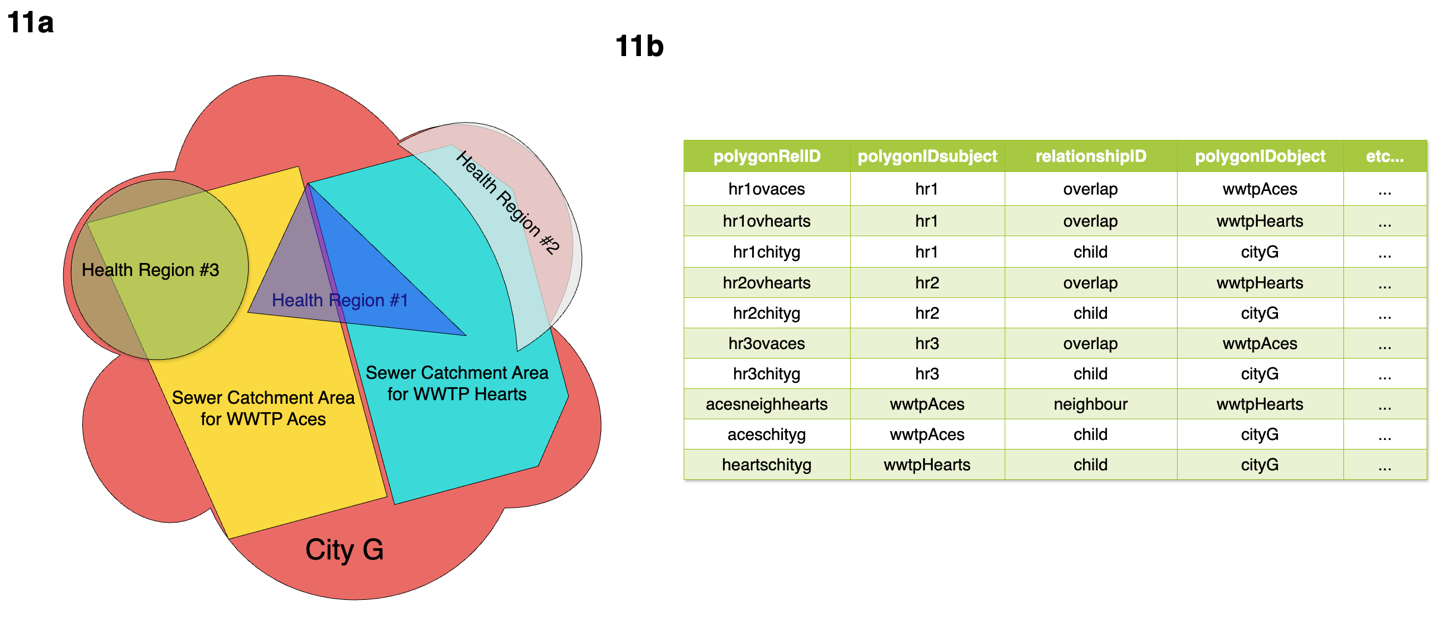}

}

\caption{\label{fig-11}Polygon relationship examples showing overlapping
and nested spatial geometries, with example data entry for the
polygonRelationships table (Figures 11a--11b).}

\end{figure}

In version 3 we have added one new relationships table, and two new
parent-type fields. The new relationship table is
``polygonRelationships'', which allows for overlapping polygons, or
nested polygons, to have that relationship made explicit. The impetus
for the addition of this table was that different agencies may use
different polygons to cover different areas. For example, a health
region may partially overlap with a wastewater catchment area, or it may
be entirely contained within it. There may even be multiple health
regions within a sewage catchment area, but they are both fully
contained within a city. An example of overlapping polygons is found in
\textbf{Figure 11a}. Making the overlapping spatial geometry more
explicit was something users said was essential for their data
management, and so the polygon relationship table was born. An example
of the polygonRelationships table being populated for the example of
City G can be found in \textbf{Figure 11b}. The polygonRelationships
table can be read in the formula: ``{[}polygonIDsubject{]} is
{[}relationshipID{]} to {[}polygonIDobject{]}''. For example, from row
one of \textbf{Figure 11b}, Health region \#1 (hr1) is overlapping WWTP
Aces' catchment area (wwtpAces). The ``polygonRelID'' is the unique
identifier for each row, and the PK for that table.

Within two of the new tables added in version 3 there are two fields
operating like parent fields to group rows together from within a table.
These two fields are the ``actionGrpID'' from the public health actions
table, and the ``pipelineID'' from the calculations table. We think a
balanced approach to intra-table grouping for defining relationships,
and external table relationship mapping for more complex cases allows
users to record complex linkages as simply as possible, preserving
detailed metadata without overburdening data generators.

\hypertarget{conclusions}{%
\section{4. Conclusions}\label{conclusions}}

WWS as it scales up globally is poised to save many lives as a valuable
tool in the global public health surveillance toolkit. The utility of
surveillance is, however, the data it generates, and that data is only
valuable insofar as it is useful and useable. The FAIR data principles,
among other data justice principles like data feminism, cognitive
interoperability, and the EOSC interoperability framework, provide a
useful metric to which WWS data can aspire. In meeting these principles,
WWS data can uphold data justice and do the greatest good for the
largest number of people. While there are many paths to be taken to
accomplish and adhere to these principles, using the PHES-ODM as a
database structure is a well-supported way to accomplish these goals.

Beyond just supporting use and re-use, as well as interoperability and
accessibility, the model also exists with various support tools to makes
its adoption and use as easy as possible for users. This includes robust
written {[}56{]} and video documentation {[}57{]}, a message board
{[}58{]}, as well as sharing {[}55{]}, validation {[}53-54{]}, and
mapping tools {[}46{]}. This means that if a user starts with the model
today, they have the information at their disposal to get started within
minutes and have the infrastructure on hand to validate and share their
data, without any investment of their own development time. The model is
entirely modular and scalable as well, so users can start with a very
basic program and data template, and the model will be there to expand
and grow with their program as it develops.

As we continue in a global age of polycrisis {[}25{]} there will be
increasing political pressure and competition for which issues to
prioritise. The COVID-19 pandemic was one arm of the polycrisis, and it
fueled a great deal of advancement in public health preparedness. Not
least of these developments was the establishment of many WWS programs
globally, and the convening of many expert research groups. Even major
global non-profit agencies, like the Bill \& Melinda Gates and the
Rockefeller Foundations have taken up the banner of WWS establishment in
low-resource settings {[}7-8{]}. As the needle shifts, however, and
different aspects of the polycrisis pull more focus, we find ourselves
at the nadir of WWS. As many smaller programs close (as almost all
provincial WWS programs in Canada have closed, transferring
responsibility to the federal program), funding priorities shift away
from health research, and public interest shifts away from surveillance,
the hard-won progress in WWS faces an existential threat.

Ensuring that the data generated by these programs is useful and of high
quality is now essential for cementing WWS programs for the future. All
surveillance programs already have data reporting standards and
dictionaries. There is often, as is the case with LOINC and SNOMED,
widespread global adoption of these dictionaries in other health
surveillance systems. This not only allows for care integration within
hospitals and health regions, but also across regions and even globally.
``Futureproofing'' WWS and its data is already a focus in several major
coalitions and action groups; there is a GLOWACON technical working
group {[}9{]}, an ELIXIR working group {[}60{]}, the WHO is working on
data-related projects and questions {[}61{]}, among many others. What is
often the focus of these groups, however, are analytical and quality
suggestions, modelling approaches, and what data to record. The last
item is one relevant to the work of the PHES-ODM, but there is a noted
lack of advice and focus on \emph{how} to record data. The field has
changed since the PHES-ODM was first developed and there are now several
data models and standards that act as strong candidates for
standardizing WWS data. The problem has instead shifted from one of a
lack of options, to one of a lack of adoption. Even with the launch of
the much-expanded version 2 of the PHES-ODM, there were already other
models and dictionaries being prepared, with different strengths and
focuses. While the PHES-ODM has focused more on PCR testing, the PHA4GE
format is very focused on sequencing data, for example (a larger
comparison summary can be found in \textbf{Table 1}). There is a gap
between the perceived need and the action on WWS data, and our hope is
that by continuing to share the PHES-ODM, its strengths, and its ease of
use, that we can help make WWS data standardisation and interoperability
a reality.

The issues facing WWS today, particularly in the realm of data, require
structural solutions. Relying on the prescience of over-burdened data
generators, or their good will and ability to go above and beyond will
not lead to success. These programs and individuals are already
overburdened, but by providing instruction, support, and tools, we can
solidify these programs, their data, and the advances made in the last 6
years for the benefit of future generations.

\textbf{Author Contributions:} Conceptualization, M.T. and D.M.;
methodology, M.T., J.D.T., N.H., J.L., and D.M.; software, M.W.;
validation, N.H., J.L., and E.S.S.; resources, P.V.R. and D.M.;
writing---original draft preparation, M.T.; writing---review and
editing, M.T., J.D.T., J.L., M.W., E.S.S., C.B., P.V.R. and D.M.;
visualization, M.T.; supervision, P.V.R and D.M.; project
administration, M.T., C.B. and D.M.; funding acquisition M.T., C.B.,
P.V.R. and D.M. All authors have read and agreed to the published
version of the manuscript.

\textbf{Funding:} Financial support for the development of the PHES-ODM
was provided by the CIHR-funded network, CoVaRR-Net (Coronavirus
Variants Rapid Response Network) {[}Funding Reference Number: 175622{]}
and the Public Health Agency of Canada. The Ontario Ministry of the
Environment, Conservation and Parks provided funding for the PHES-ODM
validation toolkit. The National Sciences and Engineering Research
Council of Canada, the Fonds de Recherche du Québec, and the
Molson-Trottier Foundation supported salaries, scholarships, travel
expenses, sample collection and laboratory analysis. Peter Vanrolleghem
holds the Canada Research Chair on Water Quality Modelling.

\textbf{Data Availability Statement:} All data and documentation
presented in this paper is openly available. Details, tables, and
resources can be found at the project website
(https://www.phes-odm.org), the GitHub repository for the project
(https://github.com/Big-Life-Lab/PHES-ODM), or the Open Science
Foundation repository (https://osf.io/ab9se/overview). Any further
queries for support can be directed to the Discourse message board
(https://odm.discourse.group), or the corresponding author.

\textbf{Acknowledgments:} As an open project, the PHES-ODM benefits from
the contribution of its steering committee, the core user group, and
many people and organisations that provide comments and suggestions.
Thank you as well to Shane Kirkham, the designer for the PHES-ODM and
other Big Life Lab projects who generates icons, graphics, and branding.
The PHES-ODM is grateful for development platforms that provide freely
available resources for open-source projects, including GitHub, the Open
Science Foundation, and Discourse.

\textbf{Conflicts of Interest:} The authors declare no conflicts of
interest

\hypertarget{abbreviations}{%
\section{Abbreviations}\label{abbreviations}}

The following abbreviations are used in this manuscript:

\begin{longtable}[]{@{}
  >{\raggedright\arraybackslash}p{(\columnwidth - 2\tabcolsep) * \real{0.5000}}
  >{\raggedright\arraybackslash}p{(\columnwidth - 2\tabcolsep) * \real{0.5000}}@{}}
\toprule\noalign{}
\begin{minipage}[b]{\linewidth}\raggedright
COD
\end{minipage} & \begin{minipage}[b]{\linewidth}\raggedright
Chemical Oxygen Demand
\end{minipage} \\
\midrule\noalign{}
\endhead
\bottomrule\noalign{}
\endlastfoot
EDGE & Enterics, Diagnostics, Genomics \& Epidemiology \\
FAIR & Findable, Accessible, Interoperable, and Reuseable \\
FK & Foreign Key \\
GIS & Geographic Information System(s) \\
GISAID & Global Initiative on Sharing All Influenza Data \\
GMA & Greater Metropolitan Area \\
LOINC & Logical Observation Identifiers Names and Codes \\
MIxS & Minimum Information about any (x) Sequence \\
NH4+-N & Nitrogen present as ammonium \\
NWSS & National Wastewater Surveillance System \\
PCR & Polymerase Chain Reaction \\
PHA4GE & Public Health Alliance for (4) Genomic Surveillance \\
PHAC & Public Health Agency of Canada \\
PHES-EF & Public Health and Environmental Surveillance Evaluation
Framework \\
PHES-ODM & Public Health and Environmental Surveillance Open Data
Model \\
PK & Primary Key \\
PMMoV & Pepper Mild Mottle Virus \\
RKI & Robert Koch Institute \\
TSS & Total Suspended Solids \\
UBA & Umweltbundesamt (German Federal Environment Agency) \\
USCDC & United States Centers for Disease Control and Prevention \\
WWE & Wastewater Epidemiology \\
WHO & World Health Organization \\
WWS & Wastewater Surveillance \\
WWTP & Wastewater Treatment Plant \\
\end{longtable}

\hypertarget{references}{%
\section{References}\label{references}}

\begin{enumerate}
\def\labelenumi{\arabic{enumi}.}
\tightlist
\item
  Singh, S., Ahmed, A. I., Almansoori, S., Alameri, S., Adlan, A.,
  Odivilas, G., Chattaway, M. A., Salem, S. B., Brudecki, G., \& Elamin,
  W. (2024). A narrative review of wastewater surveillance: pathogens of
  concern, applications, detection methods, and challenges.
  \emph{Frontiers in Public Health, 12}, 1445961.
  https://doi.org/10.3389/fpubh.2024.1445961
\item
  van der Drift, A., Welling, A., Arntzen, V., Nagelkerke, E., van der
  Beek, R., \& de Roda Husman, A. (2025). Wastewater surveillance
  studies on pathogens and their use in public health decision-making: a
  scoping review. \emph{Science of The Total Environment, 993}, 179982.
  https://doi.org/10.1016/j.scitotenv.2025.179982
\item
  Naughton, C., Roman, F., Alvarado, A. G., Tariqi, A. Q., Deeming, M.
  A., Kadonsky, K. F., Bibby, K., Bivins, A., Medema, G., Ahmed, W.,
  Katsivelis, P., Allan, V., Sinclair, R., \& Rose, J. B. (2023). Show
  us the data: global COVID-19 wastewater monitoring efforts, equity,
  and gaps. \emph{FEMS Microbes, 4}, xtad003.
  https://doi.org/10.1093/femsmc/xtad003
\item
  COVIDPoops19. (2024). \emph{ArcGIS dashboard}.
  https://www.arcgis.com/apps/dashboards/c778145ea5bb4daeb58d31afee389082
\item
  Keshaviah, A., Diamond, M. B., Wade, M. J., \& Scarpino, S. V. (2023).
  Wastewater monitoring can anchor global disease surveillance systems.
  \emph{The Lancet Global Health, 11}(6), e976--e981.
  https://doi.org/10.1016/S2214-109X(23)00170-5
\item
  Diamond, M. B., Whistler, T., Rando, K., Nwachukwu, C., \& Yousif, M.
  (2024). Policy dimensions of global wastewater surveillance.
  \emph{Bulletin of the World Health Organization, 102}(9), 622--622A.
  https://doi.org/10.2471/BLT.24.292245
\item
  Rockefeller Foundation. (n.d.). \emph{Wastewater surveillance}.
  Retrieved March 3, 2026, from
  https://www.rockefellerfoundation.org/initiatives/wastewater-surveillance/
\item
  Bill \& Melinda Gates Foundation. (n.d.). \emph{Enterics, diagnostics,
  genomics \& epidemiology (EDGE)}. Retrieved March 3, 2026, from
  https://www.gatesfoundation.org/our-work/programs/global-health/enterics-diagnostics-genomics-and-epidemiology
\item
  GLOWACON (Global Consortium for Wastewater and Environmental
  Surveillance for Public Health). (n.d.). Retrieved March 3, 2026, from
  https://glowacon.org
\item
  Manuel, D. G., Amadei, C. A., Campbell, J. R., Brault, J.-M., \&
  Veillard, J. (2022). Strengthening public health surveillance through
  wastewater testing: An essential investment for the COVID-19 pandemic
  and future health threats. \emph{World Bank}.
  https://doi.org/10.1596/36852
\item
  van Panhuis, W. G., Paul, P., Emerson, C., Grefenstette, J., Wilder,
  R., Herbst, A. J., Heymann, D., \& Burke, D. S. (2014). A systematic
  review of barriers to data sharing in public health. \emph{BMC Public
  Health, 14}, 1144. https://doi.org/10.1186/1471-2458-14-1144
\item
  Wilkinson, M., Dumontier, M., Aalbersberg, I. J., et al.~(2016). The
  FAIR guiding principles for scientific data management and
  stewardship. \emph{Scientific Data, 3}, 160018.
  https://doi.org/10.1038/sdata.2016.18
\item
  Manuel, D. G., Therrien, J.-D., Thomson, M., Sion, E.-S., Maere, T.,
  Nicolaï, N., Vanrolleghem, P. A., \& the PHES-ODM Research Group/Big
  Life Lab. (2021). \emph{PHES-ODM (Version 1.0.0)} {[}Computer
  software{]}. OSF. https://doi.org/10.17605/OSF.IO/49Z2B
\item
  Therrien, J.-D., Thomson, M., Sion, E.-S., Lee, I., Maere, T.,
  Nicolaï, N., Manuel, D. G., \& Vanrolleghem, P. A. (2024). A
  comprehensive, open-source data model for wastewater-based
  epidemiology. \emph{Water Science and Technology, 89}(1), 1--19.
  https://doi.org/10.2166/wst.2023.409
\item
  Mathieu, E., Ritchie, H., Rodés-Guirao, L., Appel, C., Gavrilov, D.,
  Giattino, C., Hasell, J., Macdonald, B., Dattani, S., Beltekian, D.,
  Ortiz-Ospina, E., \& Roser, M. (2020). COVID-19 pandemic. \emph{Our
  World in Data}. Retrieved March 3, 2026, from
  https://ourworldindata.org/coronavirus
\item
  USCDC (United States Centers for Disease Control and Prevention).
  (n.d.). National Wastewater Surveillance System (NWSS). Retrieved
  March 3, 2026, from https://www.cdc.gov/nwss/index.html
\item
  Joung, M. J., Mangat, C. S., Mejia, E., Nagasawa, A., Nichani, A.,
  Perez-Iratxeta, C., Peterson, S. W., \& Champredon, D. (2022).
  Coupling wastewater-based epidemiological surveillance and modelling
  of SARS-CoV-2/COVID-19. \emph{medRxiv}.
  https://doi.org/10.1101/2022.06.26.22276912
\item
  OClair Environnement. (2021). \emph{CETo:Connect.Predict.Prevent.}
  Retrieved March 3, 2026, from https://ceto.ca/
\item
  Shionogi \& Shimadzu. (2025). \emph{AdvanSentinel}. Retrieved March 3,
  2026, from https://advansentinel.com/en
\item
  Pepe, R. S., \& Coe, K. (2025). Data dictionaries: Essential tools for
  the ethical and transparent use of integrated data.
  \emph{International Journal of Population Data Science, 10}(2).
  https://doi.org/10.23889/ijpds.v10i2.2956
\item
  D'Ignazio, C., \& Klein, L. F. (2020). The numbers don't speak for
  themselves. In \emph{Data feminism} (pp.~36--57). MIT Press.
\item
  Regenstrief Institute. (n.d.). \emph{LOINC}. Retrieved March 3, 2026,
  from https://loinc.org/
\item
  Harrington, J. L. (2009). Why good design matters. In \emph{Relational
  database design and implementation} (3rd ed., pp.~45--50). Morgan
  Kaufmann.
\item
  Watt, A. (2014). The entity relationship data model. In \emph{Database
  Design -- 2nd Edition} (pp.~33--48). BCcampus.
\item
  Helleiner, E. (2024). Economic globalization's polycrisis.
  \emph{International Studies Quarterly, 68}(2), sqae024.
  https://doi.org/10.1093/isq/sqae024
\item
  PHES-EF. (n.d.). \emph{Public Health Environmental Surveillance
  Evaluation Framework}. Retrieved March 3, 2026, from
  https://phes-ef.org/
\item
  European Commission Joint Research Centre. (n.d.). Guidance on
  wastewater surveillance. \emph{EU Wastewater Observatory for Public
  Health}. Retrieved March 3, 2026, from
  https://wastewater-observatory.jrc.ec.europa.eu/\#/guidance/3
\item
  Esri. (n.d.). Polygon. \emph{GIS Dictionary}. Retrieved March 3, 2026,
  from https://support.esri.com/en-us/gis-dictionary/polygon
\item
  Corcho, O., Eriksson, M., Kurowski, K., Ojstersek, M., Choirat, C.,
  van de Sanden, M., \& Coppens, F. (2021). EOSC interoperability
  framework. Publications Office of the European Union.
  https://doi.org/10.2777/620649
\item
  Vogt, L. (2025). The CLEAR principle. \emph{Journal of Biomedical
  Semantics, 16}(1), 18. https://doi.org/10.1186/s13326-025-00340-7
\item
  Emerson, S. D., McLinden, T., Sereda, P., Yonkman, A. M., Trigg, J.,
  Peterson, S., Hogg, R. S., Salters, K. A., Lima, V. D., \& Barrios, R.
  (2024). Secondary use of routinely collected administrative health
  data. \emph{International Journal of Population Data Science, 9}(1),
  1--12. https://doi.org/10.23889/ijpds.v9i1.2407
\item
  Kapitan, D., Heddema, F., Dekker, A., Sieswerda, M., Verhoeff, B. J.,
  \& Berg, M. (2025). Data interoperability in context. \emph{Journal of
  Medical Internet Research, 27}, e66616. https://doi.org/10.2196/66616
\item
  Narayanan, A., Toubiana, V., Barocas, S., Nissenbaum, H., \& Boneh, D.
  (2012). A critical look at decentralized personal data architectures.
  \emph{arXiv}. https://arxiv.org/abs/1202.4503
\item
  Gomstyn, A., \& Jonker, A. (2026, February 20). What is data
  interoperability? \emph{IBM Think}. Retrieved March 3, 2026, from
  https://www.ibm.com/think/topics/data-interoperability
\item
  Christen, P., \& Schnell, R. (2023). Thirty-three myths and
  misconceptions about population data. \emph{International Journal of
  Population Data Science, 8}(1).
  https://doi.org/10.23889/ijpds.v8i1.2115
\item
  Wang, X., Williams, C., Liu, Z. H., \& Croghan, J. (2019). Big data
  management challenges in health research. \emph{Briefings in
  Bioinformatics, 20}(1), 156--167. https://doi.org/10.1093/bib/bbx086
\item
  NORMAN Network. (2020). SARS-CoV-2 in wastewater (NORMAN Database
  System). Retrieved March 3, 2026, from
  https://www.norman-network.com/nds/sars\_cov\_2/
\item
  Global Water Pathogens Project. (2020). Wastewater SPHERE. Retrieved
  March 3, 2026, from https://sphere.waterpathogens.org/
\item
  Griffiths, E. J., Timme, R. E., Mendes, C. I., Page, A. J., Alikhan,
  N.-F., Fornika, D., Maguire, F., Campos, J., Park, D., Olawoye, I. B.,
  Oluniyi, P. E., Anderson, D., Christoffels, A., Gonçalves da Silva,
  A., Cameron, R., Dooley, D., Katz, L. S., Black, A., Karsch-Mizrachi,
  I., \ldots{} MacCannell, D. R. (2022). Future-proofing and maximizing
  the utility of metadata: The PHA4GE SARS-CoV-2 contextual data
  specification package. \emph{GigaScience}, 11, giac003.
  https://doi.org/10.1093/gigascience/giac003
\item
  Paull, J. S., Barclay, C., Cameron, R., Dooley, D., Gill, I., Abraham,
  D., et al.~(2025). Fixing the plumbing: Building interoperability
  between wastewater genomic surveillance datasets and systems using the
  PHA4GE contextual data specification {[}Preprint{]}. \emph{OSF
  Preprints}. https://doi.org/10.31219/osf.io/z79vk\_v1
\item
  RKI (Robert Koch Institute). (2025). AMELAG technical guide for
  wastewater surveillance. Retrieved March 3, 2026, from
  https://www.rki.de/EN/Topics/Research-and-data/Surveillance-panel/Wastewater-surveillance/Guideline.pdf
\item
  RKI (Robert Koch Institute), \& UBA (Umweltbundesamt (German Federal
  Environment Agency)). (2026). Wastewater Surveillance AMELAG {[}Data
  set{]}. \emph{Zenodo}. https://doi.org/10.5281/zenodo.19091863
\item
  Genomic Standards Consortium. (n.d.). Minimum Information about any
  (x) Sequence (MIxS), Minimum Information about any Metagenome or
  Environmental Sequence (MIMS), Wastewater/Sludge Extension.
  MIxS:0016013 -- Wastewater surveillance environmental package.
  Retrieved March 3, 2026, from
  https://genomicsstandardsconsortium.github.io/mixs/0016013/
\item
  D'Aoust, P.M., Hegazy, N., Ramsay, N.T. et al.~SARS-CoV-2 viral titer
  measurements in Ontario, Canada wastewaters throughout the COVID-19
  pandemic. \emph{Sci Data} 11, 656 (2024).
  https://doi.org/10.1038/s41597-024-03414-w
\item
  PHES-ODM (Public Health Environmental Surveillance Open Data Model).
  (n.d.-a). Wide-names. PHES-ODM documentation. Retrieved March 11,
  2026, from https://docs.phes-odm.org/wide-names.html
\item
  Big Life Lab. (n.d.-a). PHES-ODM Mapper {[}Computer Software{]}.
  \emph{GitHub}. Retrieved March 3, 2026, from
  https://github.com/Big-Life-Lab/PHES-ODM-Mapper
\item
  Alshehri, M. (n.d.). Background. \emph{EpiWeeks documentation}.
  Retrieved March 3, 2026, from
  https://epiweeks.readthedocs.io/en/stable/background.html
\item
  Levade, I., Khan, A. I., Chowdhury, F., Calderwood, S. B., Ryan, E.
  T., Harris, J. B., LaRocque, R. C., Bhuiyan, T. R., Qadri, F., Weil,
  A. A., \& Shapiro, B. J. (2021). A combination of metagenomic and
  cultivation approaches reveals hypermutator phenotypes within Vibrio
  cholerae--infected patients. \emph{mSystems}, 6(4), e00889-21.
  https://doi.org/10.1128/mSystems.00889-21
\item
  N'Guessan, A., Tsitouras, A., Sanchez-Quete, F., Goitom, E., Reiling,
  S. J., et al.~(2022). Detection of prevalent SARS-CoV-2 variant
  lineages in wastewater and clinical sequences from cities in Québec,
  Canada {[}Preprint{]}. \emph{medRxiv}.
  https://doi.org/10.1101/2022.02.01.22270170
\item
  Hegazy, N., Peng, K. K., D'Aoust, P. M., et al.~(2025). Variability of
  clinical metrics in small population communities. \emph{ACS ES\&T
  Water, 5}(4), 1605--1619. https://doi.org/10.1021/acsestwater.4c00958
\item
  USCDC (United States Centers for Disease Control and Prevention).
  (2025, September 29). About wastewater data. Retrieved March 3, 2026,
  from https://www.cdc.gov/nwss/about-data.html
\item
  Brown, A. W., Kaiser, K. A., \& Allison, D. B. (2018). Issues with
  data and analyses. \emph{Proceedings of the National Academy of
  Sciences, 115}(11), 2563--2570.
  https://doi.org/10.1073/pnas.1708279115
\item
  PHES-ODM (Public Health Environmental Surveillance Open Data Model).
  (n.d.-b). PHES-ODM validation documentation. Retrieved March 3, 2026,
  from https://validate-docs.phes-odm.org/
\item
  PHES-ODM (Public Health Environmental Surveillance Open Data Model).
  (n.d.-c). PHES-ODM validator. Retrieved March 3, 2026, from
  https://validate.phes-odm.org/
\item
  Big Life Lab. (n.d.-b). PHES-ODM sharing library. \emph{GitHub}.
  Retrieved March 3, 2026, from
  https://github.com/Big-Life-Lab/PHES-ODM-sharing
\item
  PHES-ODM (Public Health Environmental Surveillance Open Data Model).
  (2026). PHES-ODM documentation. Retrieved March 3, 2026, from
  https://docs.phes-odm.org/
\item
  PHES-ODM (Public Health Environmental Surveillance Open Data Model).
  (n.d.-d). PHES-ODM video resources {[}Video collection{]}.
  \emph{Vimeo}. Retrieved March 3, 2026, from
  https://vimeo.com/user/126292027/folder/6496228
\item
  PHES-ODM (Public Health Environmental Surveillance Open Data Model).
  (n.d.-e). ODM discourse forum. Retrieved March 3, 2026, from
  https://odm.discourse.group/latest
\item
  Big Life Lab. (n.d.-c). PHES-ODM issues. \emph{GitHub}. Retrieved
  March 3, 2026, from https://github.com/Big-Life-Lab/PHES-ODM/issues
\item
  COVID-19 Data Portal. (n.d.). Partners and working groups. Retrieved
  March 3, 2026, from
  https://www.covid19dataportal.org/partners?activeTab=Working\%20groups
\item
  WHO (World Health Organization). (n.d.). Wastewater and environmental
  surveillance (WES). Retrieved March 3, 2026, from
  https://www.who.int/teams/environment-climate-change-and-health/water-sanitation-and-health/sanitation-safety/wastewater
\end{enumerate}

\end{document}